\documentclass[a4paper,nofootinbib]{revtex4}
\usepackage[utf8]{inputenc}
\usepackage{lipsum}
\usepackage{xcolor}
\usepackage{amssymb,amsmath,amsthm,amsfonts,mathrsfs}
\usepackage{graphicx}
\usepackage{tabularx}
\usepackage{booktabs}
\usepackage[hidelinks]{hyperref}
\usepackage[capitalise]{cleveref} 

\usepackage{xcolor}
\usepackage{subfigure}
\usepackage[normalem]{ulem}

\newcommand{\Erf}{\text{Erf}\left(\frac{\sqrt{2}}{R}r \right)}
\newcommand{\xdot}{\dot{x}}
\newcommand{\tdot}{\dot{t}}
\newcommand{\rdot}{\dot{r}}
\newcommand{\thetadot}{\dot{\theta}}
\newcommand{\phidot}{\dot{\varphi}}
\newcommand{\rLR}{r_\text{LR}}
\newcommand{\rH}{r_\text{H}}
\newcommand{\dd}{\text{d}}          % please guys, use this command for differentials as dx
\newcommand{\ii}{\text{i}}          % please guys, use this command for the imaginary unit i

\def\be#1\ee{\begin{align}#1\end{align}}
\newcommand{\lb}{\label}

\begin{document}

\title{\textbf{Semiclassical spacetimes at super-Planckian scales from delocalized sources}}

\author{A. Akil$^{a,b}$}
\email{aakil@connect.ust.hk}

\author{M. Cadoni$^{c,d}$}
\email{mariano.cadoni@ca.infn.it}

\author{L. Modesto$^{a}$}
\email{lmodesto@sustech.edu.cn}

\author{M. Oi$^{c,d}$}
\email{mauro.oi@ca.infn.it}

\author{A. P. Sanna${}^{c,d}$}
\email{asanna@dsf.unica.it}

\affiliation{$^{a}$Department of Physics, Southern University of Science and Technology, Shenzhen 518055, China}
\affiliation{$^{b}$Department of Physics, The Hong Kong University of Science and Technology, Hong Kong, China}
\affiliation{$^{c}$Dipartimento di Fisica, Universit\`a di Cagliari, Cittadella Universitaria, 09042 Monserrato, Italy}
\affiliation{$^{d}$I.N.F.N, Sezione di Cagliari, Cittadella Universitaria, 09042 Monserrato, Italy}

\begin{abstract}
We derive the gravitational field and the spacetime metric generated by sources in quantum superposition of different locations. We start by working in a Newtonian  approximation, in which the effective gravitational potential is computed as the expectation value of the gravitational potential operator in a Gaussian distribution of width $R$ for the position of the source. The effective gravitational potential is then covariantly uplifted to a fully relativistic metric in general relativity, describing the spacetime generated by averaging over the state of such sources. These results are then rederived and extended by adopting an independent construction in terms of quantum reference frames. We find three classes of quantum effective metrics which are all asymptotically flat and reproduce the Schwarzschild metric at great distances. The solutions differ, however, in the inner core. The quantum uncertainty $\Delta r\sim R$ in the position of the source prevents the radius of the transverse two-sphere to shrink to zero. Depending on the strength of the quantum superposition effects, we have either a nonsingular black hole with a ``quantum hair'' and an event horizon, a one-way wormhole with a critical null throat or a traversable wormhole. We also provide a detailed study of the geometric and thermodynamic properties of the spacetime structure for each of these three families of models, as well as their phenomenology. 
\end{abstract}

\maketitle

\tableofcontents
\section{Introduction}

The presence of singularities in general relativity (GR) \cite{Penrose:1964wq,Hawking:1970zqf,Penrose:1969pc}, both for black-holes and cosmology or even deviations from Newtonian dynamics at galactic scales or the accelerated expansion of the universe \cite{Verlinde:2016toy,Cadoni:2017evg,Cadoni:2018dnd,Tuveri:2019zor,Cadoni:2020mgb}, can be regarded as a problem that could possibly find a natural resolution in a quantum theory of gravity. 

In this paper, instead of trying to construct a fully consistent, i.e., finite or renormalizable, theory of quantum gravity (QG),  we follow a bottom-up approach, namely we start from GR and  quantum  theory and we try to gradually insert more ``quantumness'' in Gravity.

This is suited, for instance, to study quantum superpositions of reference frames or detectors \cite{Giacomini:2017zju, Foo:2020xqn, Foo:2021fno,Tjoa:2022oxv}, entanglement of quantum systems mediated by gravitational fields \cite{Bose:2017,Vedral:2017}, and the investigation of the behavior of gravitational sources (and even spacetimes) in quantum superpositions \cite{Carlesso_2019,Christodoulou:2018cmk,Foo:2020xqn, Foo:2020jmi,Foo:2021fno, Foo:2022ktn,Foo:2022dnz}. The advantage of these approaches is that they do not rely on a definite ultraviolet (UV) formulation of QG, so that they are likely to be experimentally tested in the foreseeable future. In fact, there is a number of table-top experiments, not only being designed and discussed in the literature \cite{Bose:2017,Vedral:2017, Howl:2018qdl, Howl:2020isj,Krisnanda:2019glc}, but attempts to implement them are ongoing as well \cite{Westphal:2020okx, Fein:2019, Aspelmeyer:2022fgc}. On the other hand, deviations from GR could be also probed and measured in the strong gravity regime, i.e., via gravitational-wave (GW) experiments (see, e.g., Refs.~\cite{Barack:2018yly,Cardoso:2016oxy,Carson:2019kkh,Konoplya:2016pmh,Gair:2012nm}) or, potentially, in black-hole shadow observations (see, e.g., Refs.~\cite{Lima:2021las,Younsi:2021dxe,Nampalliwar:2021oqr,EventHorizonTelescope:2020qrl}).

This bottom-up approach faces, however, a number of difficulties. 
For instance, it was argued in \cite{Penrose:1996cv,Penrose:2014nha} that gravity may cause decoherence, forcing the collapse of the quantum wave functions of matter, hence leaving no space for quantum superpositions of gravitational states. According to Penrose, the fundamental problem lies in the inconsistency between general covariance and the equivalence principle with the linearity of quantum mechanics.
This idea was further explored for  quantum states near a black hole \cite{PhysRevLett.111.021302,PhysRevLett.95.120404,PhysRevD.82.064006}. It was, however, shown in \cite{Howl:2022oqz} that, when the black hole is put in a superposition of masses, this decoherence of the quantum state largely decreases. Penrose's arguments were also challenged in \cite{Giacomini:2017zju,Giacomini:2020ahk}, where a unitary locally-inertial-reference-frame transformation was derived within a set of assumptions, supporting the validity of the equivalence principle for observers in a quantum superposition of reference frames.

The main goal of the present paper is trying to build a bridge between these different approaches. Without making any assumption about the underlying fundamental quantum theory of gravity, we derive an effective description of gravity emerging from quantum superposition of  configurations of the source. We first follow a simplified approach by working in a  Newtonian framework and regarding the standard gravitational potential as an operator acting on a Hilbert space spanned by product states of the source of the gravitational field and of a massive probe. We consider the source in a superposition of different locations in space, with a general isotropic probability amplitude $\phi(r)$. We then derive the effective Newtonian potential by taking the expectation value of the gravitational-potential operator, and the effective metric is derived using a covariant uplifting method. We then reproduce and extend the previous results by using a  more general approach.  We still consider a source in a quantum superposition of different locations, but we assume, in addition, that the metric is described by the classical Schwarzschild solution in each branch of the superposition. We then derive the explicit form of the effective metric by assuming that the source is in a quantum state described by a Gaussian wave-packet of width $R$.

The resulting spacetime metric is asymptotically flat and quickly reduces to the Schwarzschild one at large distances. However,  important differences emerge in the inner core of the solutions. The metric is invariant under reflection of the radial coordinate $r \longleftrightarrow -r$, so that it describes two asymptotically-flat and equivalent regions. Moreover, due to  the quantum uncertainty $\Delta r\sim  R$ in the position of the source, the radius of the transverse $S^2$ in the metric does not shrink to zero for $r \to 0$, but reaches a non-zero $R$-dependent value. The latter represents the radius of a throat connecting the two asymptotically-flat regions, and thus resembles the throat of a wormhole. Depending on the strength of the quantum-superposition effects, our metric describes three classes of objects: $(1)$ nonsingular black holes with an event horizon and a ``quantum hair''; $(2)$ one-way critical wormholes; $(3)$  traversable (in the sense of Morris-Thorne \cite{Morris:1988cz}) wormholes.

Our approach does not rely on a would-be specific microscopic theory of gravity. The simplest and more general guess for the effective theory is GR sourced by an anisotropic fluid. This type of fluid has been extensively used to construct stellar and black-hole models (both singular and regular) and in cosmology to address the dark matter and dark energy problems (for an incomplete list, see, e.g., Refs.~\cite{Bowers:1974tgi,cosenza1981some,Bayin:1985cd,Dehnen:2002fi,Visser:2003ge,Chirenti:2007mk,Chan:2011ayt,Aluri:2012re,Harko:2013wsa,Cadoni:2017evg,Raposo:2018rjn,Simpson:2018tsi,Simpson:2019mud,Cadoni:2020izk,Cadoni:2021zsl,Kumar:2021oxa,Cadoni:2022chn} and references therein). We therefore compute the stress-energy tensor and discuss the associated energy conditions by assuming our effective metric to represent an exact solution of Einstein's field equations, sourced by an anisotropic fluid. As expected, we find that all the usual  energy conditions are violated in all the three models. 

We then study the thermodynamic properties of the black-hole model. We find  two thermodynamic branches of black-holes: those in the ``Hawking branch'', which are unstable with respect to their radiation (they have negative specific heat), and those which instead represent thermodynamically stable configurations and have positive specific heat. By computing the free energy, we show that the latter are always thermodynamically preferred. Using the general entropy formula recently proposed in Ref.~\cite{Cadoni:2022chn} we show that the extremal black hole configuration  not only has  zero temperature, but also is a zero-entropy state. We also revisit the Hawking radiation spectrum and show it is Planckian, but with a different surface gravity. We compute the evaporation time, which turns out to be infinite in the extremal limit, thus confirming the thermodynamic stability of this configuration. 

Finally, we extensively analyze the phenomenological properties of our spacetimes, which could possibly give observable signatures in the near future. We study the geodesic structure in detail, focusing on time-like and null geodesics. In both cases, we analyze the evolution of geodesics congruence, showing that in neither of the two cases we have formation of caustics, which thus further confirms the geodesic completeness of our spacetime. This is a clear consequence of the violation of the energy conditions, which allows to circumvent Penrose's singularity theorems.  Additionally, we compute the position of the light ring, i.e., the position of the last unstable photon orbit, showing that the presence of $R$ causes potentially detectable deviations from the standard Schwarzschild prediction.

We also study scalar perturbations in this spacetime. While for small values of $R$, the effective potential in the Klein-Gordon equation has a single peak, for the stellar wormhole we  observe a double peak. This indicates the possibility of having characteristic signatures in the quasi-normal modes (QNMs) spectrum, namely echoes \cite{Cardoso:2016rao,Cardoso:2016oxy,Abedi:2016hgu,Maggio:2020jml,Maggio:2021ans, Chakraborty:2022zlq}. For models with a single peak, we exploit the construction of Ref.~\cite{Cardoso:2008bp} to derive an analytical expression of the quasi-normal frequencies in the eikonal regime. \\

The paper is organized as follows. In \cref{sec:MetricDerivation}, we derive the metric for our models following the two approaches mentioned above. 
The general geometric properties of the metric are studied in \cref{Metric}, where we also investigate the usual energy conditions and find violation of all of them. We also explicitly prove that the horizonless wormhole is traversable.  
\Cref{ThermodynamicsandHRadiation} contains an extensive analysis on the thermodynamic properties of the black-hole model, its Hawking radiation and the evaporation time. 
In \cref{sec:Phenomenology}, we investigate the phenomenological properties of our models. Specifically, we analyze the time-like and null geodesics and the evolution of their congruence and we show that the spacetime is geodesically complete. In the case of null-geodesics, we also compute the position of the light ring. In \cref{QNMsfrequency}, moreover, we compute the analytical expression of the quasi-normal frequencies in the black-hole case in the eikonal regime. 
We draw our conclusions in \cref{Conclusions}.
Throughout the entire paper, we adopt natural units, i.e., $c = \hbar = 1$.

\section{The derivation of the metric}
\label{sec:MetricDerivation}

Localized gravitational sources have been studied for centuries, and can be described through Newtonian mechanics or GR, depending on the physical settings. However, quantum mechanics showed that matter cannot be completely localized. Therefore, it is natural to study quantum delocalized gravitational sources. Here, we consider a point-like particle in a quantum superposition of different locations, and we probe its gravitational field. We will first study the quantum corrected Newtonian potential, and do a covariant uplifting to derive the spacetime metric that it generates. The resulting ``quantum'' metric will turn out to be a regular black hole that has interesting properties. We will then proceed by considering the superposition in a full covariant framework. We will use the formalism established in \cite{Giacomini:2020ahk,delaHamette:2021iwx}, through which one can construct a quantum superposition of classical spacetimes. We assume GR to hold in each branch of the superposition.
%We assume that in each branch of the superposition, GR holds.
Then, we will use the resulting state to compute the expectation value of the metric operator and get the effective quantum spacetime metric. In principle, in such a situation, the probe would get in a joint superposition (entanglement) with the source, as shown in \cite{delaHamette:2021iwx} and as intuitively expected. However, we are here only interested in an average/statistical description  of the source, which is the heart of semi-classical approximation.

\subsection{Quantum Newtonian potential uplifting}
\label{sectQNP}

In this section we will work in the framework of Newtonian gravity. We  assume that a point source of mass $M$ interacts gravitationally with a probe P of mass $m$ through the Newtonian potential. This source, however, is assumed to be in a quantum superposition of different locations. We use a spherical coordinate system and we assume that, in the classical limit, the classical source is located at the origin of the radial coordinate $r$. The Hilbert space describing the system is the cross product $\mathcal H_M \otimes \mathcal H_m$ of the Hilbert spaces $\mathcal H_M$, $\mathcal H_m$ describing separately the source and the probe, respectively. 
The gravitational-potential operator describing the system is 
\be 
\hat V= %\frac{ \hat P_{M} ^2 \otimes I_m }{2 M} + \frac{I _{M} \otimes \hat p_m ^2 }{2 m} 
- \frac{G M m  }{ |\hat r_M \otimes I_m - I_M \otimes \hat r_m| },
\ee
where $\hat r_M$, $\hat r_m,$ are the position operators for the source and the probe respectively. $I _{M}$, $I_m $ are the identity operators in $\mathcal H_M$, $\mathcal H_m$ respectively.

As an initial condition, we will assume that the particles are in a product state. The source is described by 
\be
\label{psiM}
|\psi \rangle _M = \int \dd^3 r \ \phi(\bf{r}) \, |r \rangle_M\, ,
\ee
i.e.,  we express the source state in terms of a superposition of the complete set of orthonormal eigenstates of $\hat r_M$,  being $\phi(\bf{r})$ a complex function whose modulus gives the probability amplitude for the position of the source. The state of the probe, which here is treated as a localized point particle, is simply 
\be 
| \varphi \rangle_m = | r' \rangle_m .
\ee
The quantum corrected potential is given by the expectation value of $\hat V$ with respect to the joint state $ | \psi \rangle _M \otimes | \varphi \rangle_m$, which gives
\be \lb{Potential}
\langle \hat V \rangle 
&=  - _m \! \langle  \varphi| \! \otimes     _M \! \langle \psi |   \frac{G M  }{ |\hat r_M \otimes I_m - I_M \otimes \hat r_m| }   | \psi \rangle _M \otimes | \varphi \rangle_m \\
& =   - GM \int \dd^3 r \,  \frac{|\phi(\bf{r})|^2 }{| \mathbf r - \mathbf r' | }\\
&= - 2 \pi G M \int_0 ^\pi \dd \theta \sin \theta \int _0 ^\infty \dd r \, r^2 \frac{|\phi(r)|^2 }{\sqrt{r^2 + r'^2 - 2 r r' \cos \theta }}\\
&=- \frac{2\pi G M}{r'} \int_0^\infty \dd r \ r \left(r' + r- |r-r'| \right)|\phi(r)|^2,
\ee
where the last  two equalities are valid only if we assume $\phi(r)$ to be isotropic. In the case under consideration, namely a point particle classically localized at $r=0$, this assumption is satisfied. Notice that $\langle \hat V \rangle$ does not depend on the relative phases between the states $|r\rangle_M$. The latter commonly arise due to a unitary time-evolution of the joint state of the probe, which gets entangled with the source \cite{delaHamette:2021iwx}. In this work, however, we are not interested in a fine-grained picture describing  the quantum state of one particle and its entanglement with the gravitational field, but rather in the effective description of how, on average, localized particles behave near quantum delocalized sources. Computing this expectation value between eigenstates of $\hat V$, however, erases all information regarding possible relative phases in the superposition \eqref{psiM}.

Following the standard method of covariant uplifting, one can use this potential to construct some components of the spacetime metric. The idea is that this potential can be seen as some weak field limit of a general relativistic metric, which can be guessed from the potential as
\begin{equation}\label{covuplift}
    -g_{00} = g^{-1}_{rr} \equiv f(r') = 1+2\langle \hat V(r') \rangle.
\end{equation}
Note that the conventional minus sign is already inserted in the definition of $\langle \hat V \rangle$.

Looking at \cref{psiM}, the most basic requirement we can impose on $\phi(r)$ is $L^2$-integrability (so that the state \eqref{psiM} can be correctly normalized). This requirement is sufficient to guarantee asymptotic flatness (more precisely, an asymptotic Schwarzschild form) of the resulting metric at spatial infinity, as we now show. Using \cref{Potential,covuplift}, we write the metric function in terms of the probability amplitude $\phi(r)$ of the position of the source
\begin{equation}\label{metricrprime}
    f(r') =1-\frac{4\pi GM}{r'}\int_0^{\infty} \dd r \ r \left(r' + r -|r-r'| \right)|\phi(r)|^2.
\end{equation}
To get rid of the absolute value, we separate the integral into two parts, one for $r<r'$, for which $|r-r'| = -(r-r')$, and the other for $r> r'$, for which $|r-r'| = r-r'$. Therefore, the integral in the metric function is separated accordingly  
\begin{equation}\label{integralsfr}
    \frac{1}{r'}\int_0^{\infty} \dd r \ r \left(r' + r -|r-r'| \right)|\phi(r)|^2 = %\frac{1}{r'}\int_0^{r'} \dd r \ r \left(r' + r + r-r' \right)|\phi(r)|^2 + \frac{1}{r'}\int_{r'}^{\infty} \dd r \ r \left(r' + r - r + r' \right)|\phi(r)|^2\\
     \frac{2}{r'}\int_0^{r'} \dd r \ r^2 \ |\phi(r)|^2 + 2  \int_{r'}^{\infty} \dd r \ r \ |\phi(r)|^2.
\end{equation}
%
%The other constraint comes from the $L^2$-integrability of the $\phi(r)$ distribution, which translates into requiring
$L^2$-integrability and normalization of the $\phi(r)$ distribution gives a constraint on the form of the probability amplitude, namely
\begin{equation}\label{normalizationconstraint}
    4\pi \int_0^\infty \dd r \ r^2 \ |\phi(r)|^2 =1.
\end{equation}
We will consider the behavior at asymptotic infinity first. Indeed, by taking the $r' \to \infty$ limit of \cref{integralsfr}, the second integral goes to zero, since the two integral extrema become identical. This is true as long as the integral converges, so that the $r' \to \infty$ limit and the integral commute, which is guaranteed by virtue of  \cref{normalizationconstraint}. Indeed, $\phi(r)$ is $L^2$-integrable when the minimal condition
%To \maur{satisfy} the last requirement, in fact, the minimal condition to be \maur{assured} is
%
\begin{equation}
    |\phi(r)|^2 \sim \frac{1}{r^4}+\mathcal{O}(r^{-5}), \qquad \text{for } r\to \infty
\end{equation}
is satisfied. In this case, the integral in \cref{integralsfr} reduces to
\begin{equation}\label{func}
    \int_{r'}^{\infty} \dd r \, r \, |\phi(r)|^2 \sim \int_{r'}^\infty
    \dd r \, \frac{1}{r^3} \sim \frac{1}{r'^2}\to 0 \qquad \text{for } r'\to \infty.
\end{equation}
The first integral, on the other hand, is equal to $1/4\pi$ by virtue of \cref{normalizationconstraint}. In other words, the metric reduces to $f(r') =1-\frac{4\pi G M}{r'} \frac{1}{2\pi}=1-\frac{2GM}{r'}$, which is the usual, asymptotically-flat, Schwarzschild metric.

One can also show that $L^2$-integrability of $\phi(r)$ is only a necessary condition\footnote{$L^2$-integrability alone is not sufficient to guarantee regularity of the metric. For example, a distribution like $\delta(r)/r^2$, which is $L^2$-integrable, generates the usual Schwarzschild singularity when plugged into \cref{metricrprime}.} to have a nonsingular metric, i.e., a spacetime without a central singularity at $r'=0$. However, the analysis is more involved than before. We first separate the integral in the condition \eqref{normalizationconstraint} into two parts 
\begin{equation}
    \int_0^{r'}\dd r \ r^2 \ |\phi(r)|^2 + \int_{r'}^{\infty}\dd r \ r^2 \ |\phi(r)|^2 = \frac{1}{4\pi}.
\end{equation}
Next, we use this decomposition to rewrite the right-hand side of \cref{integralsfr} as follows
\begin{equation}
    \frac{2}{r'}\int_0^{r'} \dd r \ r^2 \ |\phi(r)|^2 + 2  \int_{r'}^{\infty} \dd r \ r \ |\phi(r)|^2 = \frac{1}{2\pi r'}-\frac{2}{r'}\int_{r'}^{\infty} \dd r \ r^2 \ |\phi(r)|^2 + 2  \int_{r'}^{\infty} \dd r \ r \ |\phi(r)|^2.
\end{equation}
The second integral is well-behaved by virtue of $L^2$-integrability and, in the $r' \to 0$ limit, the divergent factor $2/r'$ in front of it cancels the other divergent term $1/2\pi r'$. The last integral, instead, can be evaluated by parts to yield
\begin{equation}
    2  \int_{r'}^{\infty} \dd r \ r \ |\phi(r)|^2 =  \left[r^2|\phi(r)|^2 \biggr|^\infty_{r'} - \int_{r'}^\infty \dd r \ r^2 \ \partial_r |\phi(r)|^2 \right].
\end{equation}
As long as we consider $\phi$ as a function and not as a distribution (thus, as long as we have smearing effects), the first ``boundary'' term will always be zero. Therefore, we see that the requirement of also $\partial_r |\phi(r)|^2$ being an $L^2$-function seems to guarantee absence of singularities for $r' \to 0$. This additional condition alone, however, is still  insufficient, as it does not automatically prevent the presence of conical singularities. The latter can be avoided if the spacetime is endowed with a throat, i.e., the angular part of the metric does not shrink to zero for $r' \to 0$ \cite{Carballo-Rubio:2019fnb}. In \cref{Metric}, we will argue that an important consequence of superposing sources in different locations, together with the related uncertainty principle,  guarantees the presence of a throat whenever $\phi(r)$ is $L^2$-integrable and sharply peaked at $r=0$. 

The specific spacetime describing the local behavior of the metric near $r' \to 0$ will, of course, strongly depend on the function of $r'$
\begin{equation}\label{Frprime}
    \mathscr{F}(r') \equiv \int_{r'}^\infty \dd r \ r^2 \ \partial_r |\phi(r)|^2\, ,
\end{equation}
and on the precise form of the angular part of the metric. This is a clear manifestation of the nonlocal nature of the quantum-mechanical approach we are using. The fact that  the angular part  of the metric is unspecified in this construction prevents us from performing  a complete analysis  of the different possibilities. 

\subsection{A simple realization: Gaussian distribution}

Our approach does not allow to determine the probability amplitude function $\phi(r)$. In fact, we are not making any assumption on the fundamental QG dynamics, which should determine $\phi$. The latter is only weakly constrained by general quantum mechanical principles. It must be $L^2$-integrable, implying that it must decrease sufficiently fast as $r \to \infty$. Moreover, the existence of a classical limit, in which the mass $M$ behaves as a point particle in the Newtonian theory (or GR), requires $|\phi|^2$ to be peaked in $r=0$. The most natural and simple candidate, respecting these and the other requirements listed in the previous subsection, is a Gaussian distribution of width $R$ centered in $r=0$. That is
\be\label{gaussian}
|\psi \rangle _M = \left(\frac{2 \sqrt 2 }{{\pi^{3/2}} R^3} \right)^\frac{1}{2} \! \int \! \dd^3 r \ e^{-\frac{r^2}{R^2}} \, |r \rangle_M\, . 
\ee
Physically, this means that we are using a wave packet  with uncertainty  $\Delta r \sim R$ as a quantum state describing  the superposition of the source location states.  The resulting momentum uncertainty reads $\Delta P \sim 1/R$. We can therefore associate to our  superposition state a De Broglie length $\lambda_\text{DB} \sim R$. As we will  show in the following sections, a comparison of $\lambda_\text{DB}$ with the gravitational (Schwarzschild) radius of the source will allow us to measure the strength of quantum effects.

Plugging \cref{gaussian} into \cref{Potential} we get 
\be\label{eq:meanV}
\langle \hat V \rangle %&&= - 4 \sqrt{\frac{2}{\pi} } \frac{ G M }{  R^3 }  \int_0 ^\pi \dd \theta \sin \theta \int _0 ^\infty \dd r \, r^2 \frac{ e^{- \frac{2 r^2 }{R^2} } } {\sqrt{r^2 + r'^2 - 2 r r' \cos \theta }} \\
= -\frac{G M} {r} \Erf.
\ee
Here and in the rest of the present work, unless otherwise specified, we have dropped the prime symbol to simplify the notation. \cref{eq:meanV} gives, upon covariant uplifting,
\be
-g_{0 0} = g^{-1}_{r r} = 1 - \frac{ 2 G M}{r} \Erf.
\label{fmetric}
\ee
As we shall see in \cref{Metric}, for particular values of the width $R$, this metric has a horizon at $\rH$ but no divergences at $r=0$, as expected. 

Interestingly, the same metric function was found also in other works dealing with non-local gravity effects and black-hole mimickers in this framework \cite{Biswas:2005qr,Biswas:2011ar,Buoninfante:2019swn,Burzilla:2020utr}. 

\subsection{A more general approach}
In the previous section we derived the expectation value of the potential operator, given a source in a quantum superposition and a localized probe. Here we will take a slightly different approach which will turn out to give the same results, but further allowing for a derivation of the angular part of the metric as well. The main difference is that we will now work in a full  covariant framework and the source is treated as being in a superposition of different locations of a given classical manifold. The mathematical formalism needed to do this was introduced in \cite{Lake:2018zeg}, and  further developed by Giacomini, Brukner and others \cite{Giacomini:2017zju,Giacomini:2020ahk,Giacomini:2021aof} in a series of papers focused on quantum observers in a superposition of different reference frames. In their approach, they also consider the possibility of having a superposition of different \textit{classical} manifolds. 

They start with the state $|\Psi^{(i)} \rangle$ describing a delocalized gravitational source in a \textit{single} manifold, labeled by a fixed index $i$, and the gravitational field associated with it
\be \lb{BrukState} 
| \Psi^{(i)} \rangle = \frac{1}{2}  %\int d^4 x' \sqrt{-g_i(x')} \psi_i(x') |x'^{(i)} \rangle _P 
\int \dd ^4 x_S \sqrt{-g^{(i)}(x_S)} \, \phi_i(x_S) |g^{(i)} (x_S-x_P) \rangle |x^{(i)}_S \rangle |x^{(i)}_P\rangle.
\ee
Here, $|x^{(i)}_S \rangle$ and $|x^{(i)}_P \rangle$ are the position eigenstates of the source and the probe respectively, while, as in the previous section, $\phi_i(x_S)$ describes the probability amplitude of the source position $x_S$ in the $i\text{-th}$ manifold, whereas $|g^{(i)} (x) \rangle$ is the state describing the spacetime metric. The factor $1/2$ is due to the symmetry under the exchange of $x_S$ and $x_P$. Moreover, $|g^{(i)} (x) \rangle$ describes a classical spacetime $\mathcal M_i$, with $i$ running through the manifolds of the superposition, i.e., $\mathcal{M}= \left\{\mathcal{M}_i \right\}_{i=1, ..., N}$. By summing over the states labelled by $i$, as well as by integrating over $x_S$, we construct a quantum superposition of classical spacetimes described by the state $|\Psi\rangle = \sum_i|\Psi^{(i)}\rangle$. We stress, again, the fact that this is not meant to represent a fully consistent second quantization of the gravitational field, but it just represents a way to build a quantum superposition of classical geometries in a first  quantization framework. Thus, summing over manifolds has not the meaning of summing over different spacetime geometries in a diffeomorphism-invariant way, but it is just a formal definition of such superposition.

Now we assume that, in each manifold $\mathcal{M}_i$, it exists a metric operator $\hat g^{(i)}_{\mu \nu}(\hat x)$ acting on the Hilbert space spanned by its eigenstates $|g^{(i)} (x) \rangle$ as
\be
\hat g^{(i)}_{\mu \nu}(\hat x) |g^{(i)} (x) \rangle = g^{(i)}_{\mu \nu}(\hat x) |g^{(i)} (x) \rangle .
\ee
The eigenvalues $g^{(i)}_{\mu \nu}(\hat x)$ are not $c$-numbers, but rather operators acting on the Hilbert space spanned by the eigenstates of the coordinates $| x \rangle$. When acting on a position eigenstate, it gives the usual spacetime metric as eigenvalues
\be 
 g^{(i)}_{\mu \nu}(\hat x) | x \rangle = g^{(i)}_{\mu \nu}( x) | x \rangle.
\ee
Describing states of the gravitational field as a quantum superposition of positions and spacetimes is not straightforward and can be controversial \cite{Penrose:1996cv}. However, there was a very nice argument presented in Ref.~\cite{delaHamette:2021iwx} supporting the validity of this construction. The authors start with a massive object (the source) in a superposition of 2 locations, and a localized probe falling through. They do not construct an {\it{a priori}} superposition of spacetimes, like $|g_1\rangle + |g_2 \rangle$. Instead, they construct a quantum-reference-frame transformation which makes the source localized and leaves the free-falling probe in a superposition of 2 locations. In that case, the physics is described by the semiclassical approach. They then evolve the superposed probe state on the determined curved background, and, at the end of the evolution, they transform back to the original frame in which the source is in a superposition. The result turns out to be in exact accordance with the case where the whole process is done with the source being in a superposition of the two locations, described by the state $|g_1\rangle + |g_2 \rangle $.

The state $|\Psi \rangle$ defined above can now be used to compute the metric operator expectation value
\be
 \langle \hat g_{\mu\nu}(\hat x)\rangle \equiv\langle \Psi | \hat g_{\mu \nu}(\hat x) | \Psi \rangle 
 &= \frac{1}{4} \sum_{i,j=1}^N \int \dd^4 x_S' \sqrt{-g^{(j)}(x_S')} \phi_{j}^{\ast}(x_S') \langle g^{(j)} (x_S' - x_P') |\langle x'^{(j)}_S |  \langle x'^{(j)}_P | \hat g^{(i)}_{\mu \nu}(\hat x) \nonumber \\
 & \phantom{\frac{1}{4} \sum_{i,j=1}^N \int} \times \int \dd ^4 x_S \sqrt{-g^{(i)}(x_S)}\phi_{i}(x_S) |g^{(i)} (x_S - x_P)\rangle |x^{(i)}_S \rangle |x^{(i)}_P\rangle \nonumber\\
 &= \sum_{i=1}^{N} \int \dd^4x_S \ \sqrt{-g^{(i)}(x_S)} \ |\phi_{i}(x_S)|^2 \, \, g^{(i)}_{\mu \nu} (x_S-x_P),\lb{expMetric}
\ee
where we assumed that the metric and the position eigenstates of both the probe and the source are orthogonal to each other, and specifically (see also Refs.~\cite{Giacomini:2020ahk,Giacomini:2021aof})
\begin{equation}
\frac{1}{4}\langle g^{(j)}|g^{(i)}\rangle \langle x_S'^{(j)}|x_S^{(i)}\rangle = \frac{\delta^{(4)}\left(x_S-x_S' \right)}{\sqrt{-g^{(i)}(x_S)}} \delta^{ij}\, ,
\end{equation}
where the Kronecker delta reinforces the fact that gravitational fields on different manifolds are perfectly distinguishable. On a curved background, the distribution $|\phi_i(x)|^2$ now satisfies the normalization condition $\int \dd^4 x \ \sqrt{-g^{(i)}(x)}\ |\phi_i(x)|^2=1$.
Note also that we are summing over different manifolds in order to account for extra physical parameters which can as well be in quantum superpositions. That sum can also be an integral for continuous parameters. The mass of the source or the probe being in a superposition is a simple example (see the end of the present Section).

As previously noticed, summing over geometries, even in a first quantization framework, is a quite involved procedure. The completeness of the Hilbert space spanned by these geometries and diffeomorphism invariance are important issues one should address before performing the summing \cite{Giddings:2005id}. In order to avoid these problems and to keep things as simple as possible, we just consider superposition of the positions of a gravitational source in a single given geometry and in a given coordinate system.
We fix therefore a particular gauge, requiring that, the spacetime metric is the Schwarzschild metric. We also fix the parametrization by writing the latter in the Eddington-Finkelstein coordinates $(v, r, \theta, \varphi)$ for simplicity. Therefore, \cref{expMetric} becomes
\be 
\langle \hat g_{\mu \nu}(\hat x) \rangle = \mathcal N \, ^2 \frac{2\sqrt {2} }{ \pi^\frac{3}{2} R^3}  \int \dd v \, \dd r \, \dd \theta \, \dd \varphi \, r^2 \sin\theta \, e^{-2 r^2/R^2} g_{\mu \nu} (r-r_P).
\ee
Given that the metric is static, i.e., invariant under $v$-translations in each branch of the superposition, we have to renormalize the integral over $v$ with a renormalization-factor $\mathcal{N}^2$. Moreover, given the spherical symmetry of the metrics $g^{(i)}$, we can also integrate over $\varphi$. This yields
\be 
\langle \hat g_{\mu \nu}(\hat x) \rangle = \frac{4\pi\sqrt {2} }{ \pi^\frac{3}{2} R^3} \int \dd r \, \dd \theta \, r^2 \sin \! \theta \, e^{ -2 r^2/R^2} g_{\mu \nu} (r-r_P).
\ee
Plugging the explicit expressions of $g_{\mu\nu}$ in terms of the Schwarzschild metric, the expectation value of the metric operator reads
 \be \langle \dd s^2 \rangle = \left[ -1 + \frac{2GM}{r}\Erf \right]  \dd v^2 + 2 \dd v \dd r + \left( r^2 + \frac{3R^2}{4} \right) \dd \Omega^2, \quad \dd\Omega^2 = \dd\theta^2 + \sin^2 \theta \dd\varphi^2,
 \ee
where now the radial coordinate $r$ corresponds to the distance of the probe to the source.
Interestingly, thus, for a Gaussian distribution of the probability amplitude of the source, one gets the same metric components computed in the previous section, supporting the result, and, in addition, interesting radial components emerge.

Transforming to the Schwarzschild coordinates yields
\begin{equation}\label{MetricBrukner}
  \langle \dd s^2 \rangle = \left[ -1 + \frac{2GM}{r} \Erf \right] \dd t^2 +\frac{1}{ 1 - \frac{2GM}{r} \Erf } \dd r^2 + \left( r^2 + \frac{3R^2}{4} \right) \dd \Omega^2 .
\end{equation}
Note that we can also introduce a superposition of masses. In a simplified formulation, one can promote the mass in the Schwarzschild metric to an operator, and let it act on a state vector accounting for the dependence of the system from its ADM mass. In this way, we can consider a quantum superposition of mass eigenstates. Note that the mass operator corresponds to an observable in quantum gravity, since it is an explicitly gauge-invariant quantity. Its nonlocal nature is here inherited from the superposition of the different eigenstates.
We start therefore by writing the state as 
\be 
    | \Psi \rangle = \frac{1}{2}  %\int d^4 x' \sqrt{-g_i(x')} \psi_i(x') |x'^{(i)} \rangle _P 
    \int \dd ^4 x_S \sqrt{-g(x_S)} \,  \phi(x_S) |g(x_S-x_P)\rangle |x_S \rangle |x_P   \rangle \int \dd M \,  \psi(M)|M \rangle .
\ee
where $\psi(M)$ describes the distribution of different masses. The previously defined metric operator, on the other hand, will be promoted to $\hat g_{\mu \nu} (\hat x , \hat M)$. Then, assuming the Schwarzschild metric in each branch of the superposition, and focusing on the mass state, we have
\be  \lb{masssuper}
    \langle g_{\mu \nu} (\hat x, \hat M) \rangle = \int \dd M' \, \psi^*(M') \, \langle M' | \, \hat g_{\mu \nu} (\hat x , \hat M) \int \dd M \, \psi(M)  \, |M \rangle.
\ee
Substituting the zeroth component of the metric as an example reads
\be 
\langle g_{0 0 } \rangle = 
  \int \dd M' \, \psi^*(M') \, \langle M' |  \, \left( 1 - \frac{ 2G \hat M}{\hat r}  \right) \int \dd M \, \psi(M)  \, |M \rangle.
 \ee
Assuming $\psi(M)$ to be normalized to $1$, we have 
 \be 
 & \int \dd M \, | \psi(M) |^2 = 1  \\
& \int \dd M \, | \psi(M) |^2 M = \langle M \rangle \equiv M_{\rm cl},
 \ee
where we have identified the classical mass $M_{\rm cl}$ with the expectation value of the operator $\hat M$. When plugged together into \cref{masssuper}, one easily finds
\be  \lb{mtrcop}
 \langle g_{\mu \nu} (\hat x, \hat M) \rangle = 1 - \frac{2G M_{\rm cl} }{\hat r}.
 \ee
 The same works for the other components of the metric tensor. An important remark is that these results are totally independent of the details of the superposition, i.e., of the explicit form of the distribution $\psi(M)$. \cref{MetricBrukner} is again recovered when we compute the expectation value of the operator \eqref{mtrcop} with the position states of the source.
 
 \section{Metric Structure}
\lb{Metric} 

\begin{figure}
\centering
\subfigure[]{\includegraphics[width=0.45\textwidth]{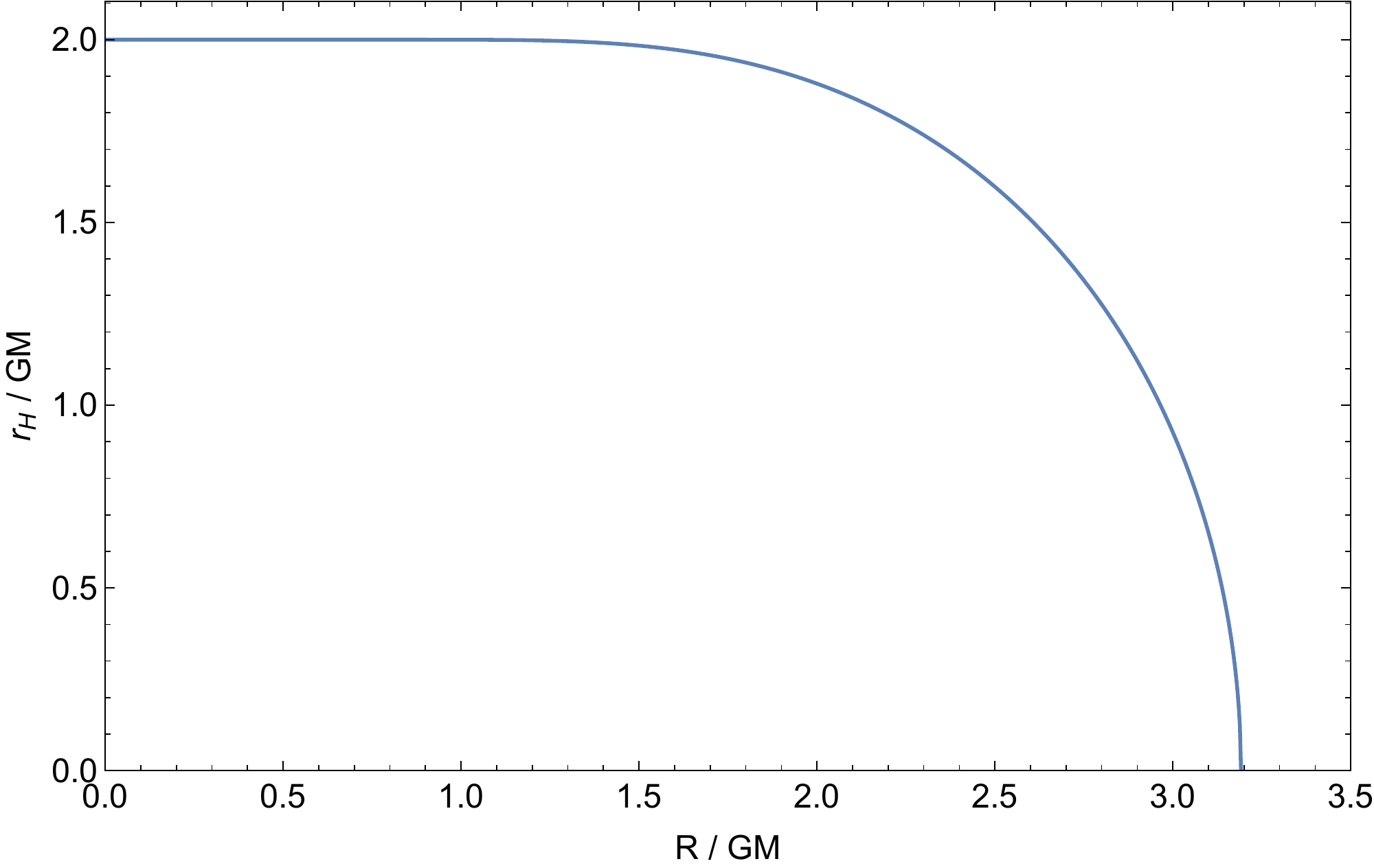}}
\hspace{0.55 cm}
\subfigure[]{\includegraphics[width=0.45\textwidth]{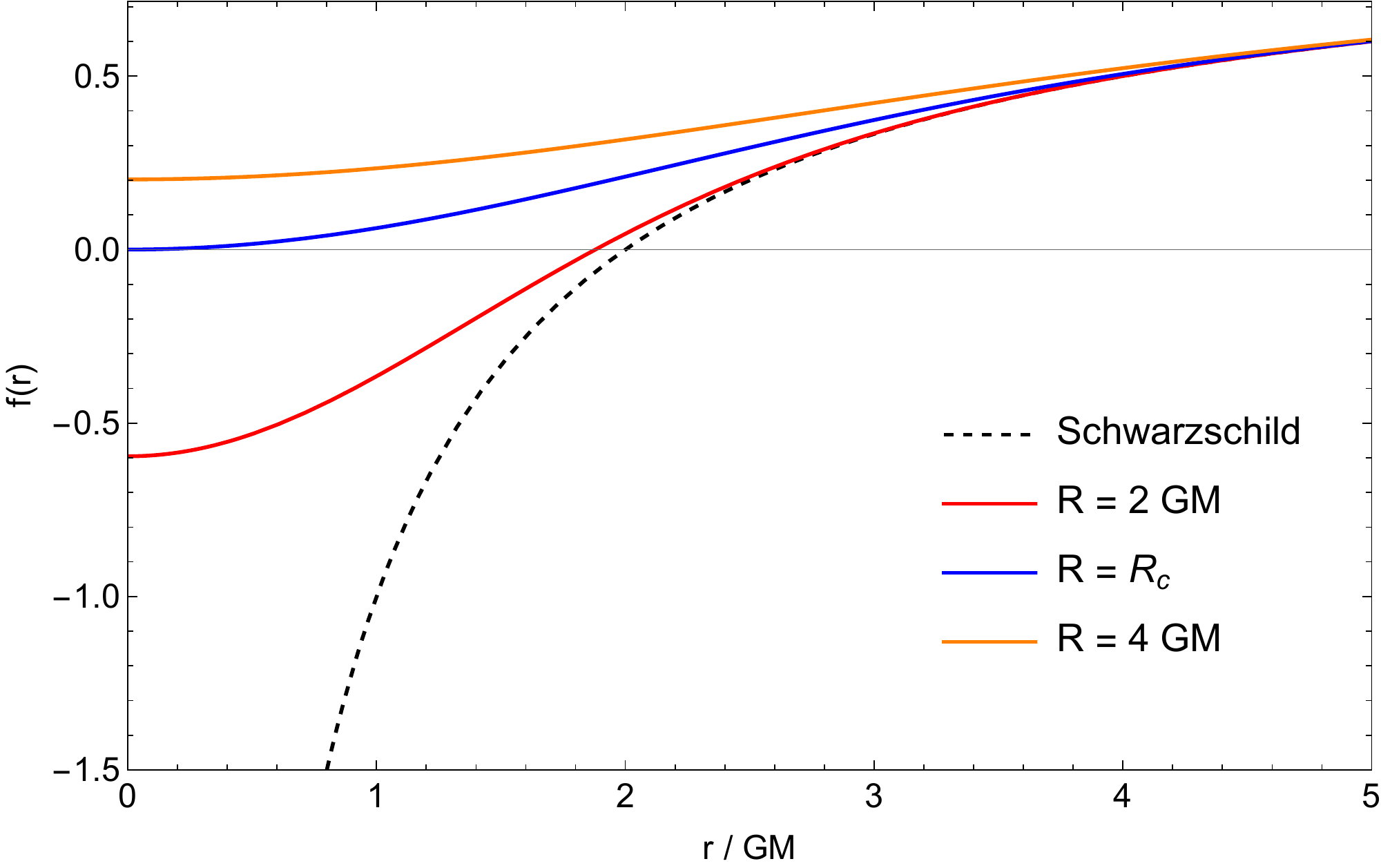}}
\caption{{\bf Left figure:} Horizon radius as a function of the smearing parameter $R$, both in units of $GM$. We see that, for values of $R$ greater than $R_\text{c} \simeq 3.19 \, GM$, the horizon disappears and we are left with a horizonless object. The horizontal line corresponds to the position of the classical Schwarzschild horizon.\\
{\bf Right figure:} Behavior of the metric function \eqref{metricf} as a function of the radial coordinate for different value of $R$: $R = 2 GM$ (solid red line), $R = R_\text{c}$ (solid blue line) and $R = 4GM$ (solid orange line). The first case corresponds to a solution with an event horizon (regular black holes), the second to a ``critical'' wormhole with a null throat, while the latter to a horizonless object, which is a two-way wormhole.}
\label{MetricGeometry}
\end{figure}

In the remainder of the paper, we will use $f(r)$ to indicate the metric function in \cref{MetricBrukner}, i.e.,
\begin{equation}\label{metricf}
    f(r) = 1-\frac{2GM}{r}\Erf.
\end{equation}
The metric reduces to the standard Schwarzschild metric for $r\gg R$. This last result can also be obtained in the $R\to 0$ limit, i.e., in the limit in which the width of the Gaussian position distribution goes to zero, which yields the standard Dirac-delta distribution, thus recovering the classical central singularity.

In the $r\to 0$ limit, instead, the metric function behaves as
\begin{equation}
\label{nearzero}
    f(r) \simeq 1-\frac{4GM\sqrt{2/\pi}}{R} + \frac{8GM\sqrt{2/\pi}}{3R^3} r^2 + \mathcal{O}(r^3).
\end{equation}
We see that there are no spacetime singularities at $r=0$. This suggests the relevance of the present approach to the construction of nonsingular black-hole models, which have recently gained increasing attention (see, e.g., Refs.~\cite{Carballo-Rubio:2019fnb,Simpson:2019mud,Carballo-Rubio:2019nel,Cadoni:2022chn}).
The local $r=0$ behavior of the metric function $f$ is similar, except for the constant term, to that of the anti-de-Sitter case. We have explicitly computed the curvature invariants for our spacetime metric (see \cref{sec:curvatureinv}) and showed that they remain finite at $r=0$. One can also easily show that the $r={\rm constant}$-time slices of our spacetime have a surface with area $\mathcal{A}=4\pi \left(r^2 + \frac{3}{4}R^2 \right)$, which is minimized at $r=0$. The radius of the two-sphere does not shrink to zero, but to the minimal non-vanishing value $\sqrt{3/4}\, R$. This means that, near $r=0$, the $t=\rm constant$ sections of our spacetime exhibit a $\mathbb{R}\times S^2$ local topology. Additionally, we have an invariance of the metric under $r \longleftrightarrow -r$. Altogether this means that the metric \eqref{MetricBrukner} describes two asymptotically-flat equivalent regions, connected through a long throat of minimal radius $\sqrt{3/4}\, R$, i.e., a wormhole. Indeed, in the $M \to 0$ limit, our metric reduces to the standard Morris-Thorne wormhole \cite{Morris:1988cz,Morris:1988tu}
\begin{equation}\label{MTWormhole}
    \dd s^2 =-\dd t^2 + \dd r^2 + \left(r^2 + \frac{3R^2}{4} \right)\dd \Omega^2.
\end{equation}
The $r \to 0$ behavior of the metric is what distinguishes our solution from other ``quantum-inspired'' regular models \cite{Hayward:2005gi,Nicolini:2005vd,Modesto:2010uh,Giugno:2017xtl,Simpson:2019mud,Casadio:2021eio,Cadoni:2022chn}. Similarly to our case, these solutions are parametrized by a quantum hair $R$ and are usually endowed with a de Sitter core, which determines the presence of two (or, more generally, an even number of) horizons. Inspection of \cref{nearzero} reveals that $f(0)$ changes sign at the critical value of $R=R_{\rm c} = 4 \sqrt{2/\pi}\, GM\simeq  3.19 \, GM$. This signalizes the presence of horizons, whose position can be easily found by computing the zeroes of $f(r)$.
For $R< R_{\rm c}$, we have one horizon, while for $R> R_{\rm c}$ we have no horizons (see \cref{MetricGeometry}). At $R = R_{\rm c}$, instead, the metric function has a zero at $r=0$ and we have an ``extremal configuration'', separating solutions with and without horizons. 

The occurrence of different solutions for different values of the parameter $R$ has a nice explanation in terms of the strength of quantum effects characterizing our quantum superposition of spacetimes. $R$ and $R_{\rm c}$ are of the order of magnitude of the De Broglie length $\lambda_\text{DB}$ of our quantum state and of the classical gravitational radius of the source $R_\text{S}=2GM$, respectively. Thus, $R\ll R_{\rm c}$ means that quantum effects are completely negligible and we are describing the classical limit of a fully localized source. Correspondingly, the solutions of the effective theory are indistinguishable from the classical Schwarzschild black hole with its singularity at $r=0$. When $R\sim R_{\rm c}$, instead, quantum effects become relevant and the solution of the gravitational theory is a ``quantum-deformed'' Schwarzschild black hole: $R$ plays the role of a quantum hair and the classical singularity at $r=0$ is resolved. Finally, $R > R_{\rm c}$ corresponds to a regime that is fully dominated by the quantum effects generated by the superposition of the source location states. On the effective gravitational theory, we have now a horizonless wormhole solution. This is a quite intriguing result, reminiscent of the $\text{ER}=\text{EPR}$ conjecture \cite{VanRaamsdonk:2010pw,Maldacena:2013xja}. When quantum effects become fully dominant, both the singularity and the horizon disappear, leaving behind a fully regular traversable wormhole. 

One could ask whether the presence of a wormhole in the effective theory is generic or a consequence of assuming the Gaussian form~\eqref{gaussian} for the distribution $\phi(r)$. We can easily show that a wormhole solution will always be present, regardless of the specific form of $\phi(r)$, whenever the latter is $L^2$-integrable and sharply peaked at $r=0$ (as required by a meaningful quantum picture and for consistency with the classical description in terms of localized source-particle) and whenever the metric is Schwarzschild in every branch of the superposition. 
Indeed, from \cref{expMetric} we see that the $g_{\theta\theta}$ component of the effective metric can be written as
\be
    \langle g_{\theta\theta}\rangle = 2\pi \int_0^\pi \dd\theta\int_0^\infty \dd r_{\rm S} \, r_{\rm S}^2 \, \sin\theta \, \left|\phi(r_{\rm S})\right|^2 (r^2+r_{\rm S}^2-2 r r_{\rm S}\cos\theta),
\ee
where $r_{\rm S}$ indicates the radial coordinate of the source. The integral over $\theta$ can be done immediately, giving
\be
    \langle g_{\theta\theta}\rangle = 4\pi \int_0^\infty \dd r_{\rm S} \, r_{\rm S}^2 \, \left| \phi(r_{\rm S}) \right|^2 (r^2+r_{\rm S}^2) = %
    4\pi r^2 \int_0^\infty \dd r_{\rm S} \,  r_{\rm S}^2 \,  \left| \phi(_{\rm S}) \right|^2 + 4\pi\int_0^\infty \dd r_{\rm S} \, r_{\rm S}^4 \, \left| \phi(r_{\rm S}) \right|^2.
\ee
Using \cref{normalizationconstraint}, the first integral gives $1/4\pi$. By defining a new dimensionless variable $\xi = r_{\rm S} / R$, the second one, can be written in the form,
\be\label{eq:proofOfSigmaForm}
    \int_0^\infty \dd r_{\rm S} \  r_{\rm S}^4 \left| \phi(_{\rm S}) \right|^2 = R^5 \int_0^\infty \dd \xi \, \xi^4 \left| \phi(\xi) \right|^2.
\ee
If we assume the integral to be convergent, as it is the case for a sharply peaked function, simple counting of dimensions in the normalization of $\phi$, gives $\phi\sim R^{-3/2}$.
This implies in turn that integral \eqref{eq:proofOfSigmaForm} gives $a^2 R^2$, where $a^2$ is some real constant. Therefore, the metric can always be written in the form
\be\label{form1}
    \dd s^2 = -f(r, R) \dd t^2 + \frac{\dd r^2}{f(r,R)} + (r^2 + a^2 R^2) \dd\Omega^2.
\ee
%In fact, applying \cref{expMetric} to a generic distribution function $\phi(r)$, which is $L^2$-integrable and has a sharp peak at $r=0$ with dispersion $\sim R$, we get a metric of the form
%
% \begin{equation}\label{form1}
%     \dd s^2 = -f(r, R) \dd t^2 + \frac{\dd r^2}{f(r,R)} + \Sigma(r,R) \dd\Omega^2,
% \end{equation}
%
%The $L^2$-integrability condition implies $\Sigma$ has the form $\Sigma=r^2+ a^2 R^2$, with $a$ a real dimensionless constant.
Additionally, if $\phi$ has a narrow maximum at $r=0$, the first integral in \cref{func} is dominated by the contribution near this maximum, so that this integral, and hence the metric function $f(r)$ in \cref{form1}, are even functions of $r$. Finally, $L^2$-integrability guarantees that the metric is asymptotically flat, as shown in \cref{sectQNP}. Altogether, these features tell us that the metric \eqref{form1} represents a wormhole.

\subsection{Effective theory and energy conditions}
\label{EnergyConditions}

From Birkhoff's theorem, the only static, vacuum-solution of Einstein's field equations is the Schwarzschild metric. Therefore, our spacetime must be sourced by some non-zero stress-energy tensor. As previously stated, we are not making any assumption about the fundamental quantum theory of gravity underlying our quantum description of spacetime. Our goal is restricted to deriving the effective description of gravity emerging from quantum superposition of positions of the source. Owing to our lack of knowledge about the underlying theory of QG, the simplest, and more general, guess on the emerging effective theory is that of GR sourced by an anisotropic fluid \cite{Bowers:1974tgi,cosenza1981some}, which is characterized by profiles for the energy density $\epsilon$ and for the radial and transverse  components of the fluid pressure, respectively given by  $p_\parallel$ and $p_\perp$. This means that the effect of the quantum superposition of spacetimes allows for an effective classical description in terms of an anisotropic fluid. This kind of fluids are very promising for parametrizing QG effects both for black holes/compact objects \cite{Dehnen:2002fi,Visser:2003ge,Hayward:2005gi,Nicolini:2005vd,Chirenti:2007mk,Chan:2011ayt,Raposo:2018rjn,Simpson:2019mud,Casadio:2021eio,Kumar:2021oxa,Cadoni:2022chn} and for galactic dynamics and cosmology \cite{Bayin:1985cd,Aluri:2012re,Harko:2013wsa,Cadoni:2017evg,Tuveri:2019zor,Cadoni:2020izk,Cadoni:2021zsl}. The information about the effective theory will be encoded in the profile $\epsilon(r)$ and the equation of state $p_\parallel=p_\parallel(\epsilon)$, whereas $p_\perp$ is determined by the conservation equation for the stress energy tensor. Assuming our metric being an exact solution of Einstein's equations, we can compute the explicit expressions of the density and the pressure components for the anisotropic fluid
\begin{subequations}
\be
    &\epsilon = \frac{-3R^3+4e^{-2r^2/R^2} G M \sqrt{\frac{2}{\pi}} \left(3R^2 + 4r^2 \right) +\frac{6 G M R^3}{r}\Erf }{2\pi G R\left(3R^2 + 4r^2 \right)^2}\, ; \label{density}\\
    &p_\parallel = \frac{-3R^3-4e^{-2r^2/R^2} G M \sqrt{\frac{2}{\pi}} \left(3R^2 + 4r^2 \right) +\frac{6 G M R^3}{r}\Erf }{2\pi G R\left(3R^2 + 4r^2 \right)^2}\, ; \label{pparallel}\\
    &p_\perp = \frac{6 R^5 r^3 + 2e^{-2r^2/R^2} G M \sqrt{\frac{2}{\pi}} r \left(9R^6 + 30 R^4 r^2 + 48 R^2 r^4 + 32 r^6 \right)-3 G M R^5 \left(3R^2 + 8r^2 \right)\Erf}{4\pi G R^3 r^3 \left(3R^2 + 4r^2 \right)^2}\, . \label{pperp}
\ee
\end{subequations}
We now analyze the energy conditions, focusing on the Null Energy Condition (NEC). In order this condition to be satisfied, we have to require both $\epsilon + p_\parallel \geq 0$ and $\epsilon + p_\perp \geq 0$ to hold globally. It is sufficient to consider that, from \cref{density,pparallel}, it follows
\begin{equation}
    \epsilon + p_\parallel = -\frac{3R^2 f(r)} {G \pi  \left(3R^2 + 4r^2 \right)^2}\, ,
\label{NECconditionexterior}
\end{equation}
with $f(r)$ given by \cref{metricf}. For $R> R_{\rm c}$, i.e., for horizonless objects, the NEC is always violated, since $f(r)>0$ everywhere (see \cref{MetricGeometry}). This means that the wormhole is potentially traversable \cite{Morris:1988cz,Visser:1995cc} (see \cref{traversability}). For $R = R_{\rm c}$, the NEC is always violated except from the point $r=0$. This means that this model represents a one-way wormhole with a null throat at $r=0$, which poses restrictions to its traversability as we will see in detail  in the following section.

In the black-hole case, things are a little more subtle. In the exterior region, the NEC is always violated, since $f(r) > 0$. In the interior, we have that the time and the radial coordinates swap, so that we now have $\epsilon=-T^r_r$ and $p_\parallel = T^0_0$, while $p_\perp$ remains unchanged. Therefore, the right-hand side of \eqref{NECconditionexterior} changes sign. This implies that also in the interior, $f(r)<0$, and thus the NEC is violated. 

For $R \to 0$ (the limit in which our spacetime reduces to the standard Schwarzschild solution), we have that $\epsilon + p_\parallel=0$ and the NEC is of course satisfied.
\begin{figure}
\centering
\subfigure[ $R = 2 G M$]{\includegraphics[width=5cm]{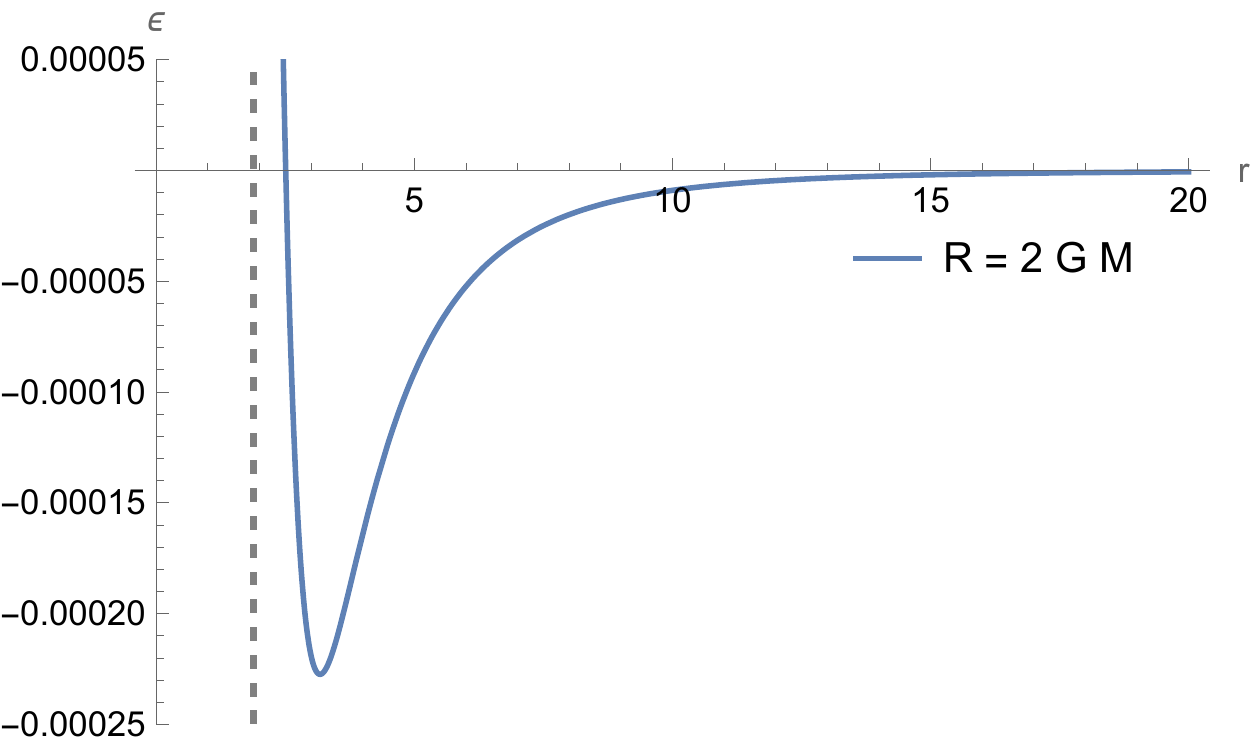}}
\hspace{0.3 cm}
\subfigure[ $R = R_{\rm c}$]{\includegraphics[width=5cm]{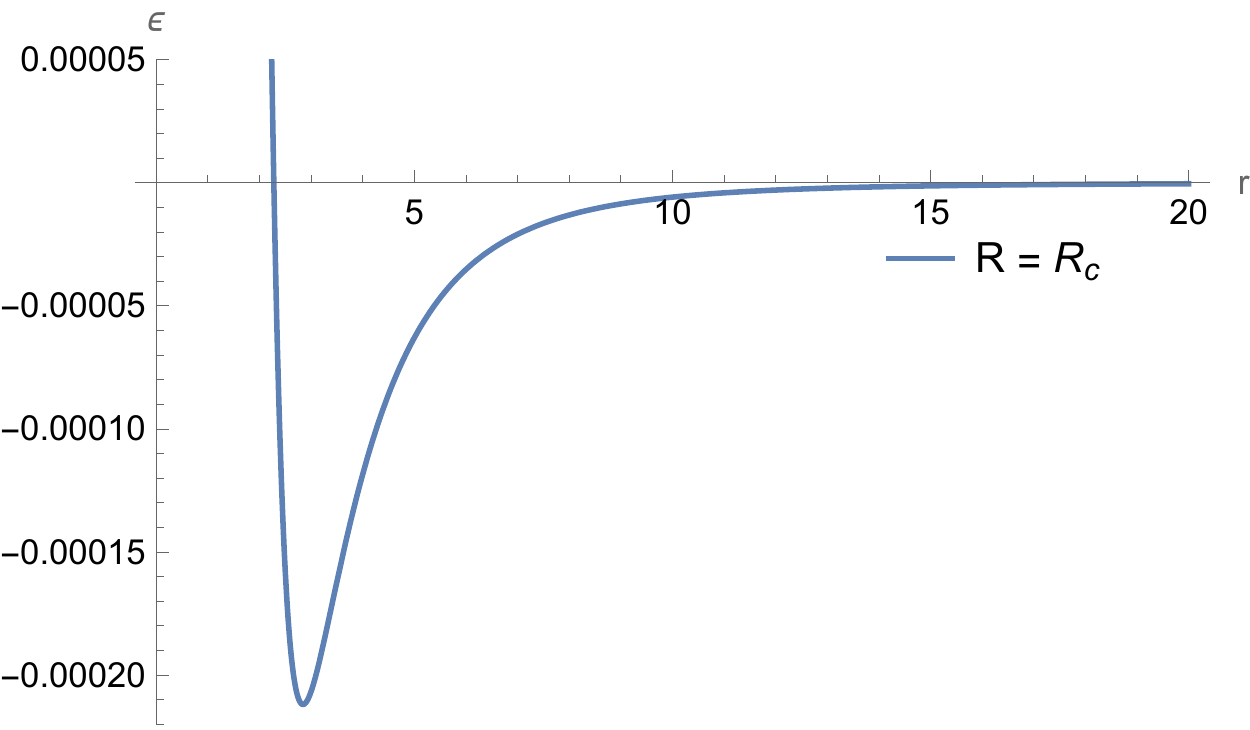}}
\hspace{0.3 cm}
\centering
\subfigure[ $R = 4 G M$]{\includegraphics[width=5cm]{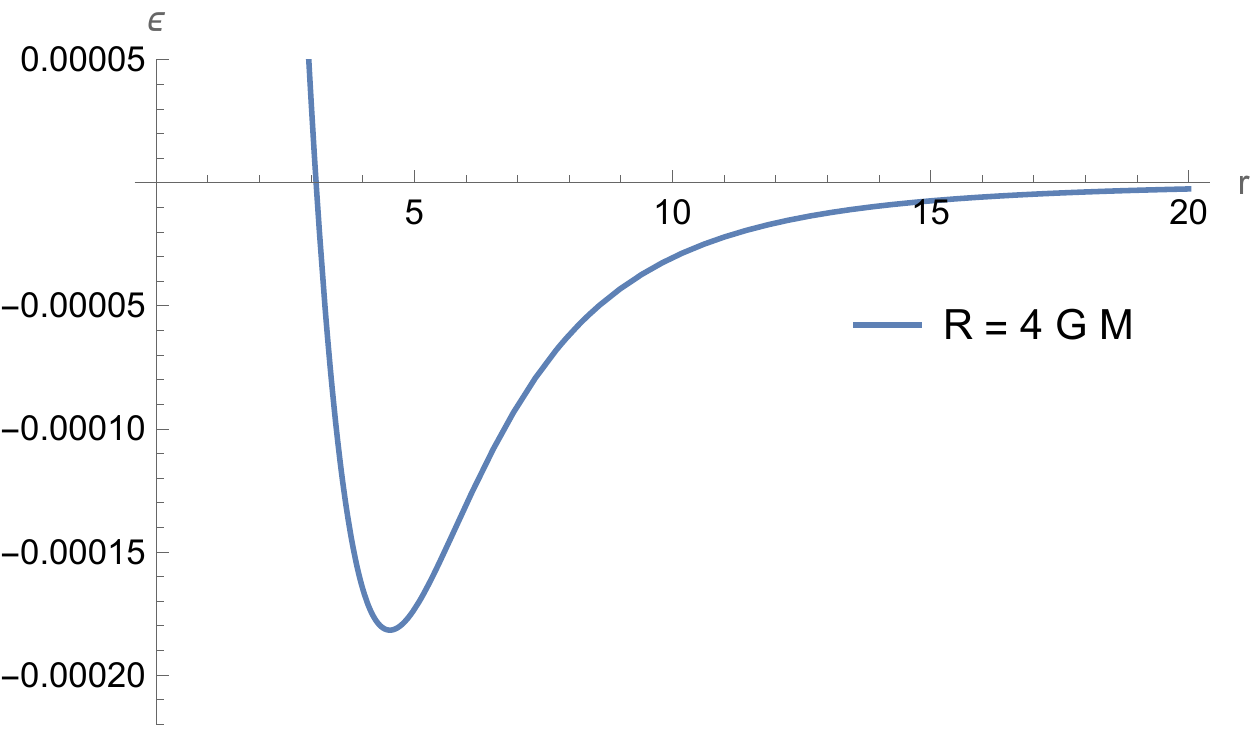}}
\caption{Density $\epsilon$ as a function of $r$ in the three cases, black-hole (case a), ``critical wormhole'' (case b) and wormhole (case c). In the first case, the dashed vertical line corresponds to the position of the event horizon. For all figures, we set $G = M = 1$.}
\label{WEC}
\end{figure}
Violation of the NEC is a sufficient condition for violating all the other energy conditions \cite{Visser:1995cc,Lobo:2020ffi}. Indeed, it has been proved \cite{Lobo:2020ffi} that, for a general metric of the form $\dd s^2 = -f(r) \dd t^2 + f^{-1}(r) \dd r^2 + \Sigma(r) \dd \Omega^2$, there is a violation of all  energy conditions (regardless of whether $t$ is a temporal or a spatial coordinate, i.e., regardless of whether we are inside or outside the horizon) whenever $f(r) \neq 0$ and $\Sigma(r)$ is non-zero everywhere and satisfies $\Sigma(r) > 0$ and $\Sigma''(r) > 0$, which is indeed the case here.

We have explicitly checked that the other energy conditions are also violated. It is worth noting that the weak energy condition $\epsilon \geq 0$ is typically strongly violated in the region near the Schwarzschild radius, whereas it holds both inside the latter and in the asymptotic ($r\to\infty$) region (see \cref{WEC}).

\subsection{Wormhole traversability}
\label{traversability}

The violation of the standard energy conditions is only a necessary, but not sufficient, condition to have an ``in-principle'' traversable wormhole \cite{Hochberg:1998ii}. An additional condition, commonly referred to as ``flaring-out'', needs to be satisfied. To properly explain the physical meaning and implications of this requirement, we write a general-wormhole metric in the standard form 
\begin{equation}\label{standardwomrholemetric}
    \dd s^2 = -e^{2\Phi(r)} \dd t^2 + \frac{\dd r^2}{1-\frac{b(r)}{r}}+r^2 \dd \Omega^2
\end{equation}
where $\Phi(r)$ and $b(r)$ are functions of $r$ only. $b(r)$ controls the spatial shape of the wormhole and is therefore called the ``shape function'', while $\Phi(r)$ is the ``redshift function''. 

The ``flaring-out'' condition guarantees that the throat does not close and remains hypothetically stable. This is achieved by requiring the induced spatial hypersurfaces on both sides of the throat to be strictly increasing with the distance from the throat itself \cite{Morris:1988cz,Visser:1995cc,Kundu:2021nwp}. To do so, we compute the proper radial distance from the throat in the wormhole spacetime \eqref{standardwomrholemetric}, which is
\begin{equation}
    \mathcal{L}(r) = \pm \int_{r_{\rm throat}}^{r}\frac{\dd r}{\sqrt{1-\frac{b(r)}{r}}}.
\end{equation}
The radius of the throat $r_{\rm throat}$ is given by the minimum of $r\left(\mathcal{L} \right)$, which translates to imposing
\begin{equation}
    \frac{\dd r}{\dd\mathcal{L}}= \pm \sqrt{1-\frac{b(r_{\rm throat})}{r_{\rm throat}}}=0,
\end{equation}
which gives $r_{\rm throat}$ as the solution of $b(r_{\rm throat})= r_{\rm throat}$. Finally, in order the proper distance to be strictly increasing on both sides of the minimum $r_{\rm throat}$, we require
\begin{equation}
    \frac{\dd^2 r}{\dd \mathcal{L}^2} = \frac{1}{2r_{\rm throat}} \left(-b'(r_{\rm throat})+\frac{b(r_{\rm throat})}{r_{\rm throat}} \right)>0.
\end{equation}
Since $b(r_{\rm throat})= r_{\rm throat}$, the ``flaring-out'' condition translates to requiring $b'(r_{\rm throat})<1$.

To explicitly analyze this condition in our model, we need first to recast our metric \eqref{MetricBrukner} into the form of \cref{standardwomrholemetric}. This is simply realized by the coordinate change $r' \equiv \sqrt{r^2 + \frac{3R^2}{4}}$, and the metric \eqref{MetricBrukner} becomes
\begin{equation}
\begin{split}
\label{BruknerWormHole}
\dd s^2 = -\left[1-\frac{2GM}{\sqrt{r'^2 - \frac{3R^2}{4}}}\text{Erf} \left(\frac{\sqrt{2}}{R} \sqrt{r'^2 - \frac{3R^2}{4}} \right)\right] \dd t^2 + \frac{r'^2 \dd r'^2}{\left(r'^2-\frac{3R^2}{4} \right)\left[1-\frac{2GM}{\sqrt{r'^2 - \frac{3R^2}{4}}}\text{Erf} \left(\frac{\sqrt{2}}{R} \sqrt{r'^2 - \frac{3R^2}{4}} \right)\right]} + r'^2 \dd \Omega^2,
\end{split}
\end{equation}
from which we immediately read the ``redshift'' and the ``shape'' functions
\begin{subequations}
\be
    &\Phi(r') =\frac{1}{2} \ln\left[1-\frac{2GM}{\sqrt{r'^2 - \frac{3R^2}{4}}}\text{Erf} \left(\frac{\sqrt{2}}{R} \sqrt{r'^2 - \frac{3R^2}{4}} \right)\right] ; \label{redshiftfunction}\\
    &b(r') =r'-\frac{1}{r'}\left(r'^2-\frac{3R^2}{4} \right)\left[1-\frac{2GM}{\sqrt{r'^2 - \frac{3R^2}{4}}}\text{Erf} \left(\frac{\sqrt{2}}{R} \sqrt{r'^2 - \frac{3R^2}{4}} \right)\right]\label{shapefunction}.
\ee
\end{subequations}
The position of the throat is given by solving the equation $b(r_\text{throat})=r_\text{throat}$, so that
\begin{equation}\label{throatposition}
    \left(r_\text{throat}^2-\frac{3R^2}{4} \right)\left[1-\frac{2GM}{\sqrt{r_\text{throat}^2 - \frac{3R^2}{4}}}\text{Erf} \left(\frac{\sqrt{2}}{R} \sqrt{r_\text{throat}^2 - \frac{3R^2}{4}} \right)\right]=0.
\end{equation}
Regularity of the redshift function \eqref{redshiftfunction} \textit{everywhere} \cite{Morris:1988cz,Visser:1995cc}, required to have traversability, implies that the quantity in square brackets in \cref{throatposition} is different from zero, which isolates the throat radius $r_\text{throat}=\sqrt{3/4}\, R$, as expected.

Taking the derivative of $b(r)$ with respect to $r'$ and evaluating it at $r'=\sqrt{3/4}\, R$ yields
\begin{equation}
    b'\left(\sqrt{\frac{3}{4}}R \right)=-1+\frac{8GM}{R}\sqrt{\frac{2}{\pi}}.
\end{equation}
For $b'(\sqrt{3/4}\, R)<1$, i.e., for $R > 4\sqrt{2/\pi}\, GM$, we have a traversable wormhole, while it is non-traversable otherwise. It is interesting to note that the same value of $R$ discriminating between traversable and non-traversable wormholes is the same discriminating between the presence or absence of an event horizon. Specifically, horizonless wormholes will be traversable, while those with an event horizon will not. The object with $R = R_{\rm c}$ falls in this last category as a particular configuration with a null throat.

This is consistent with the usual Misner and Thorne's requirement \cite{Morris:1988cz} of the absence of event horizons to guarantee traversability, as the presence of an horizon prevents two-way travel through the wormhole. This is of course a consequence of the requirement of the regularity of the redshift function \eqref{redshiftfunction}, which implies the absence of an horizon. Indeed, if we had fixed $r_{\rm throat}$ as the zero of the square bracket in \cref{throatposition}, we would have had $\Phi(r) \to -\infty$ and thus an horizon (since we would have had $e^{2\Phi}\to 0$ in \cref{standardwomrholemetric}).

\section{Thermodynamics and Hawking evaporation}
\label{ThermodynamicsandHRadiation}

\subsection{Thermodynamic properties}
\label{sec:Thermodynamics}

\begin{figure}
\centering
\includegraphics[width= 8 cm, height = 8 cm,keepaspectratio]{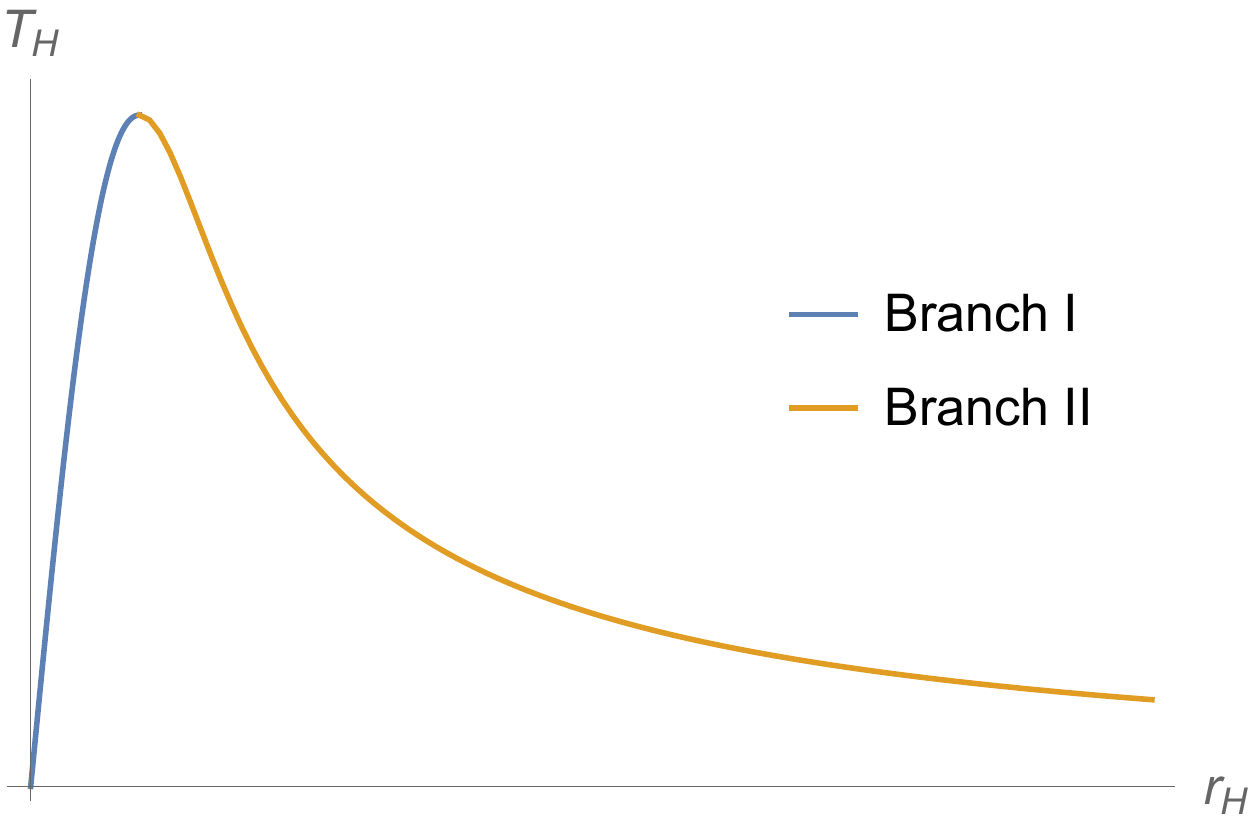}
\caption{Qualitative behavior of the temperature as a function of the event horizon radius. We highlighted the two thermodynamic branches: thermodynamic stable configurations (blue line) and the unstable (Hawking) branch (orange line).}
	\label{Fig.Temperature}
\end{figure}

From the metric function \eqref{metricf}, using the standard black-hole thermodynamic relations, we can compute both the black-hole mass and the Hawking temperature $T_\text{H}$ as functions of the event horizon radius $\rH$ and the position uncertainty $R$
\begin{subequations}
\begin{align}
    &M(\rH,R)= 
    \frac{\rH}{2G \text{Erf}\left(\frac{\sqrt{2}}{R}\rH \right)}, \label{ADMmass}\\
    &T_\text{H}(\rH,R) = \frac{1}{4\pi} \frac{\dd f(r)}{\dd r}\biggr|_{r = \rH}= \frac{G M(\rH,R)}{2\pi^{3/2} \rH^2}\left[-\frac{2\sqrt{2} \  e^{-2\rH^2/R^2}}{R}\rH + \sqrt{\pi}\  \text{Erf}\left(\frac{\sqrt{2}}{R}\rH \right) \right]. \label{Temperature}
\end{align}
\end{subequations}
Plugging \cref{ADMmass} into \cref{Temperature} yields the explicit expression of the temperature
\begin{equation}\label{Temperaturefinal}
    T_\text{H}(\rH,R) = \frac{1}{4\pi \rH} -\frac{\sqrt{\frac{2}{\pi}}\ e^{-2\rH^2/R^2}}{2\pi R \ \text{Erf}\left(\frac{\sqrt{2}}{R}\rH \right)}\, .
\end{equation}
The first term corresponds to the standard Hawking result. Indeed, it is easy to see that, in the $R \to 0$ limit, $\rH \to 2 G M$ and $T_\text{H} \to 1/8\pi G M$. In the $\rH\to 0$ limit, instead, the temperature goes as $T_\text{H}\simeq \rH/3\pi R^2 + \mathcal{O}(\rH^2)$, so it goes to zero linearly. The temperature also vanishes  as $\rH\to \infty$. This signals the non-monotonic behavior of the temperature, which must have at least an extremum somewhere. Indeed, solving $\dd T_{\rm H}/\dd \rH=0$ yields the position of the maximum $r_{\rm H, max} \simeq 0.97 \, R$. A qualitative plot of the temperature is shown in \cref{Fig.Temperature}. 

As expected, the standard thermodynamic divergence of the temperature at $\rH \to 0$ of the Schwarzschild black hole is cured. The $\rH=0$ configuration corresponds to the ``extremal'' wormhole, which, therefore, is a perfectly regular, zero-temperature state. In the limit $\rH \to 0$, $M(\rH, R) \to M_{\rm c} \equiv \frac{1}{4 G}\sqrt{\frac{\pi }{2}} \, R$, which is a non-zero value. This signals the  transition from an object with an event horizon to a horizonless one.

\begin{figure}
\centering
\includegraphics[width= 9 cm, height = 9 cm,keepaspectratio]{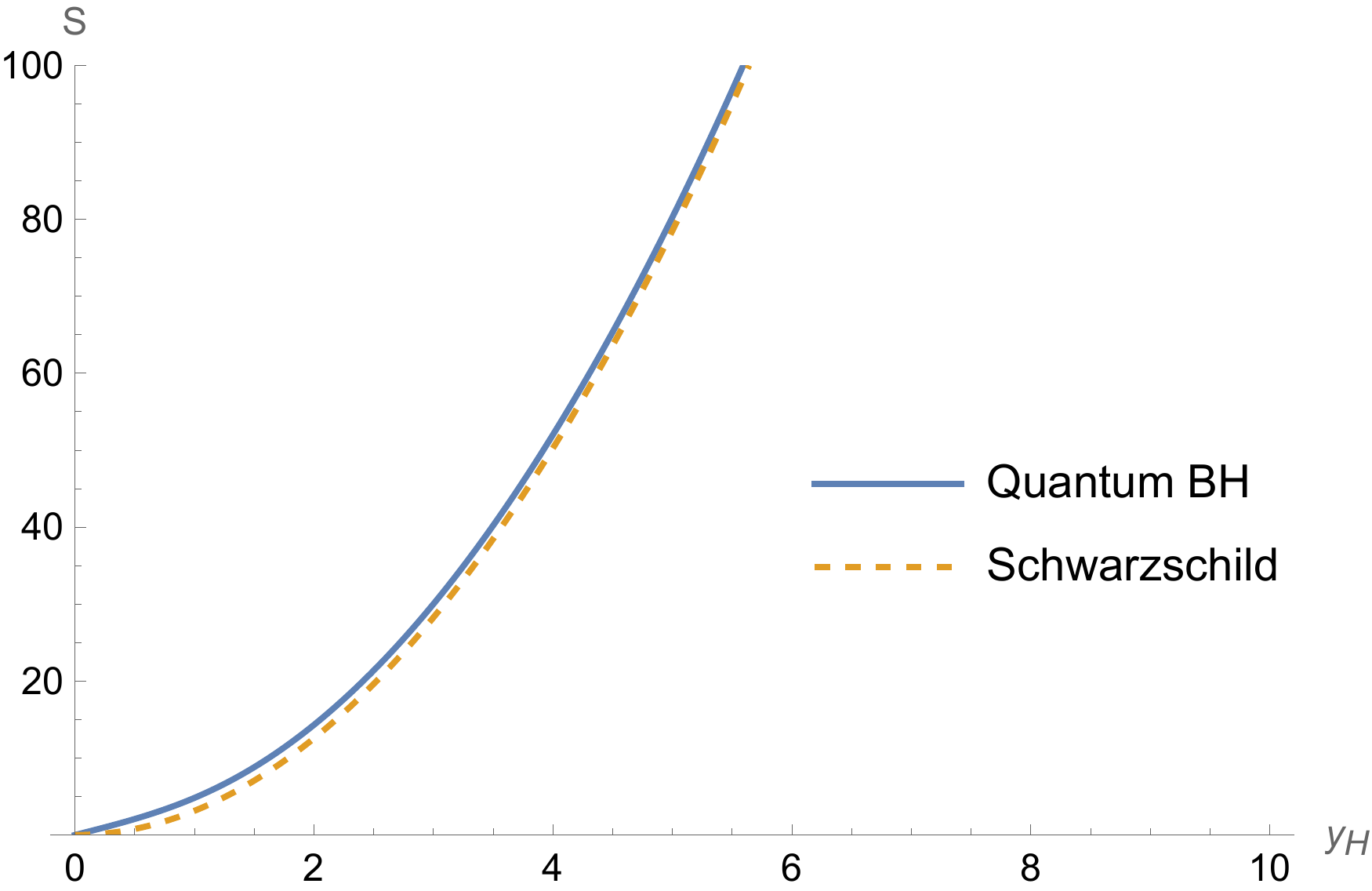}
\caption{Numerical evaluation of the entropy of the quantum black hole (solid blue line), compared to Hawking's standard result $S_{\rm H} = \mathcal{A}_\text{H}/4G$ (with $\mathcal{A}_\text{H}$ the area of the event horizon), as a function of the adimensional event horizon radius $y_\text{H}\equiv\rH/R$. We set $G=1$.}
	\label{Entropy}
\end{figure}

An important remark is that $R$ has to be considered as a quantum deformation parameter that, contrary to $M$, is not associated with conserved charges defined at infinity. This makes our quantum black hole solution drastically different from other two-parameter classes of solutions, like, e.g., the charged Reissner-Nordstr\"om solution, for which \textit{both} parameters are associated with thermodynamic potentials. Owing to this feature, we expect a first law of thermodynamics of the form $\dd M = T_\text{H} \dd S$, where $S$ is the black-hole entropy. It is known that the presence of a quantum deformation parameter $R$, not associated with a thermodynamic potential, implies violation of  the area-law for the entropy \cite{Cadoni:2022chn}. An entropy formula, which generalises the area-law and applies to  ``quantum-deformed'' black holes, has been proposed in Ref.~\cite{Cadoni:2022chn}
\begin{equation}\label{Entropyfinal}
    S = 4\pi \int_{r_{\rm min}}^{\rH} M(\rH') \ \dd \rH'.
\end{equation}
where $r_{\rm min}$ is the minimum value  attained by $\rH$.
One can easily check that \cref{ADMmass,Temperature} imply the validity of the relation $\dd M=4\pi  M T_\text{H}\dd\rH$, from which it follows that the entropy \eqref{Entropyfinal} satisfies the first principle $\dd M = T_\text{H} \dd S$.
In the case under consideration, $r_{\rm min}=0$ and, therefore, the entropy of the extremal, $T_\text{H}=0$, configuration vanishes. This $T_\text{H}=S=0$, $M\neq 0$ extremal state separates  solutions with horizons from horizonless wormholes.
In \cref{Entropy} we plotted the result of the numerical integration of the entropy expression \eqref{Entropyfinal}.

As mentioned before, the entropy formula \eqref{Entropyfinal} is a consequence of the validity of the first law of black-hole thermodynamics in its standard formulation, i.e., with $M(r)$ identified as the internal energy of the system. This eventually led to a deviations from the entropy area law. It is worth stressing that a parallel, but conceptually different, thermodynamic description can be given \cite{Ma:2014qma}, in which instead the area-law is satisfied, but the first law gets modified: the internal energy is not identified with $M$ anymore, but also the matter-fields contribution is taken into account. This is due to the extra dependence on $M$ contained in the stress-energy tensor, which leads to a first law of the form
\begin{equation}
    C(\rH, M) \, \dd M = T_\text{H}\,  \dd \left(\frac{\mathcal{A}_\text{H}}{4} \right), \qquad C(\rH, M) \equiv 1 + 4\pi \int_{\rH}^\infty \dd r \, r^2 \, \frac{\partial T^0_0}{\partial M}\, ,
\end{equation}
where $\mathcal{A}_\text{H}$ is the area of the event horizon. In the standard case, $\partial T^0_0/\partial M = 0$, $C(\rH, M) = 1$ and we recover the usual formulation of the first law. 

However, this discussion is limited to an analysis of the equations of motion and a Lagrangian description of these models is clearly required to have a precise thermodynamic interpretation of the internal energy of the system and of the entropy (see Ref. \cite{Visser:1993qa}), which would allow one to prefer one approach over the other. 

Let us end this section by briefly discussing the behavior of our solutions near the ``extremal'' configuration, i.e., the configuration with $\rH=0$, $T_{\rm H} = 0$, to gain some insights into the transition between the black-hole and the horizonless wormhole models. Expanding around this critical value, at leading order we get, for the mass and the temperature
\begin{subequations}
\be
    &M \simeq M_{\rm c} + \beta \rH^2;\\
    &T_{\rm H}\simeq \gamma \rH,
\ee
\end{subequations}
where we have defined $\beta \equiv \frac{1}{2}\frac{\dd^2 M}{\dd \rH^2}\biggr|_{\rH = 0}$ and $\gamma \equiv \frac{\dd T_{\rm H}}{\dd \rH}\biggr|_{\rH = 0}$. Combining the two expressions together, we find the scaling of the mass above extremality in terms of the temperature
\begin{equation}\label{Massgapaboveextremality}
    M-M_{\rm c} \sim \frac{\beta}{\gamma^2} T_{\rm H}^2.
\end{equation}
This scaling of the mass  above extremality with the temperature squared is typical of several black-hole models \cite{Almheiri:2014cka,Almheiri:2016fws}.

\subsection{Second-order phase-transition and free energy}

The non-monotonic behavior of the temperature \eqref{Temperaturefinal} depicted in \cref{Fig.Temperature} signalizes the presence of a non-trivial thermodynamic phase portrait and of a second-order phase transition occurring  when the temperature  reaches the  maximum, for the solutions with $0\le R\le R_\text{c}$. To see this, we consider the specific heat of the solution, given by
\begin{equation}\label{sh}
C= \frac{\dd M}{\dd T} =\frac{\dd M}{\dd\rH} \left(\frac{\dd T}{\dd\rH} \right)^{-1}.
\end{equation}
Being $\dd M/\dd\rH$ always positive, the non-monotonic behavior of $T_\text{H}$ implies that 
\begin{itemize}
    \item For $r_{\rm c}< \rH<r_{\text{H, max}}$, $\dd T/\dd \rH$ is positive and thus $C>0$;
    \item For $\rH>r_{\text{H, max}}$, $\dd T/\dd \rH$ is negative and thus $C<0$; 
    \item For $\rH=r_{\text{H, max}}, \,\dd T/\dd \rH=0$  and thus $C\to \infty$. 
\end{itemize}

\begin{figure}
\centering
\includegraphics[width= 9 cm, height = 9 cm,keepaspectratio]{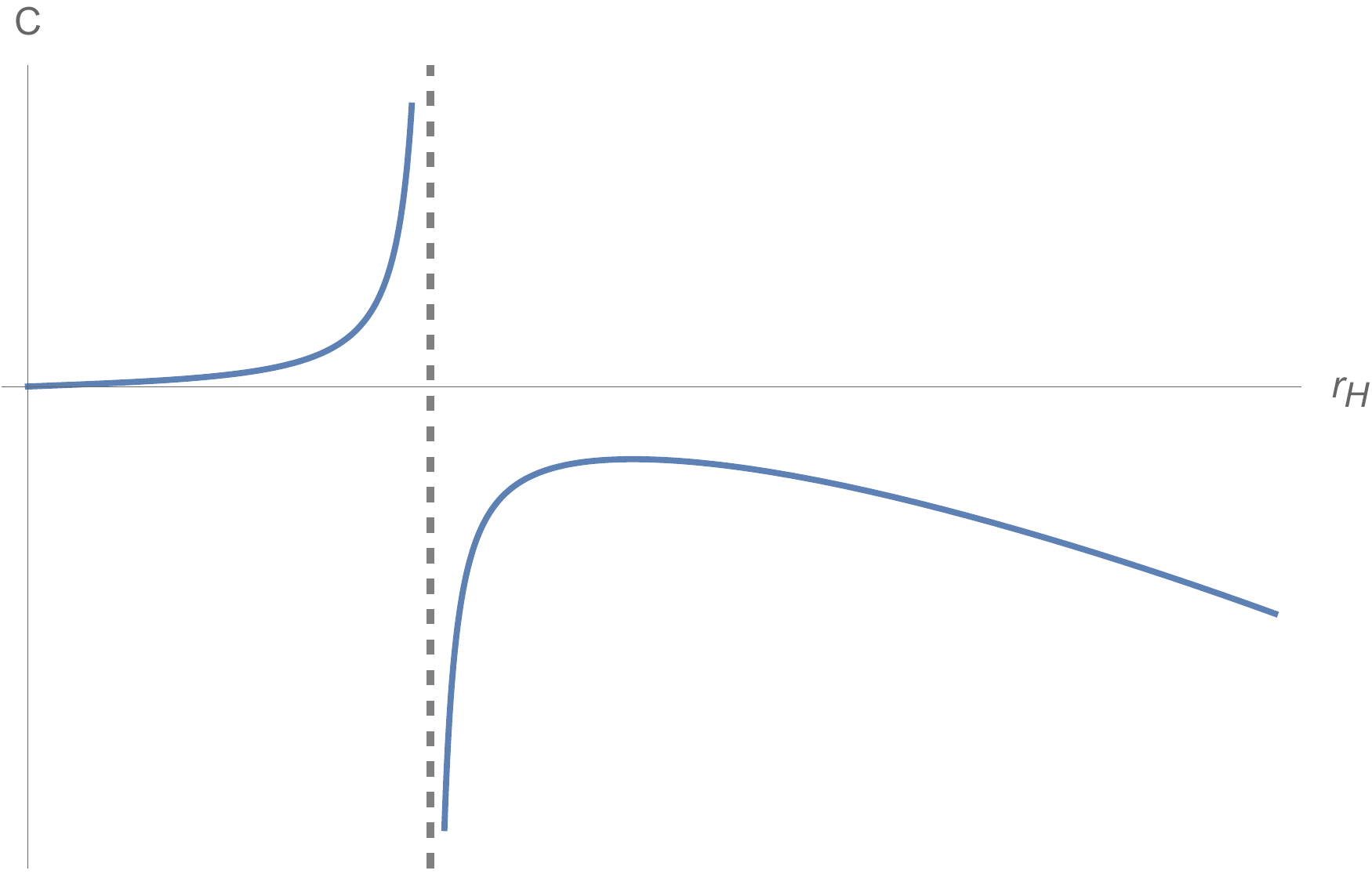}
\caption{Qualitative behavior of the specific heat \eqref{sh} as a function of the horizon radius $\rH$. The dashed vertical line corresponds to the position of the maximum of the temperature \eqref{Temperaturefinal}.}
	\label{SpecificHeat}
\end{figure}

The qualitative behavior of the specific heat is represented in \cref{SpecificHeat}. 
The second-order phase transition distinguishes between two thermodynamic branches. Branch $I$, corresponding to the left side of the temperature in \cref{Fig.Temperature}, pertains to black holes with positive specific heat, which therefore can be considered at equilibrium with their radiation. 
Branch $II$, instead, describes black holes with large event-horizon radii (right part of the temperature \eqref{Temperature} in \cref{Fig.Temperature}), which are instead characterized by a negative specific heat and therefore are unstable with respect to their radiation. This branch corresponds to ``classical'' black holes, for which the effects of the smearing parameter $R$ are negligible.    

\begin{figure}
\centering
\includegraphics[width= 9 cm, height = 9 cm,keepaspectratio]{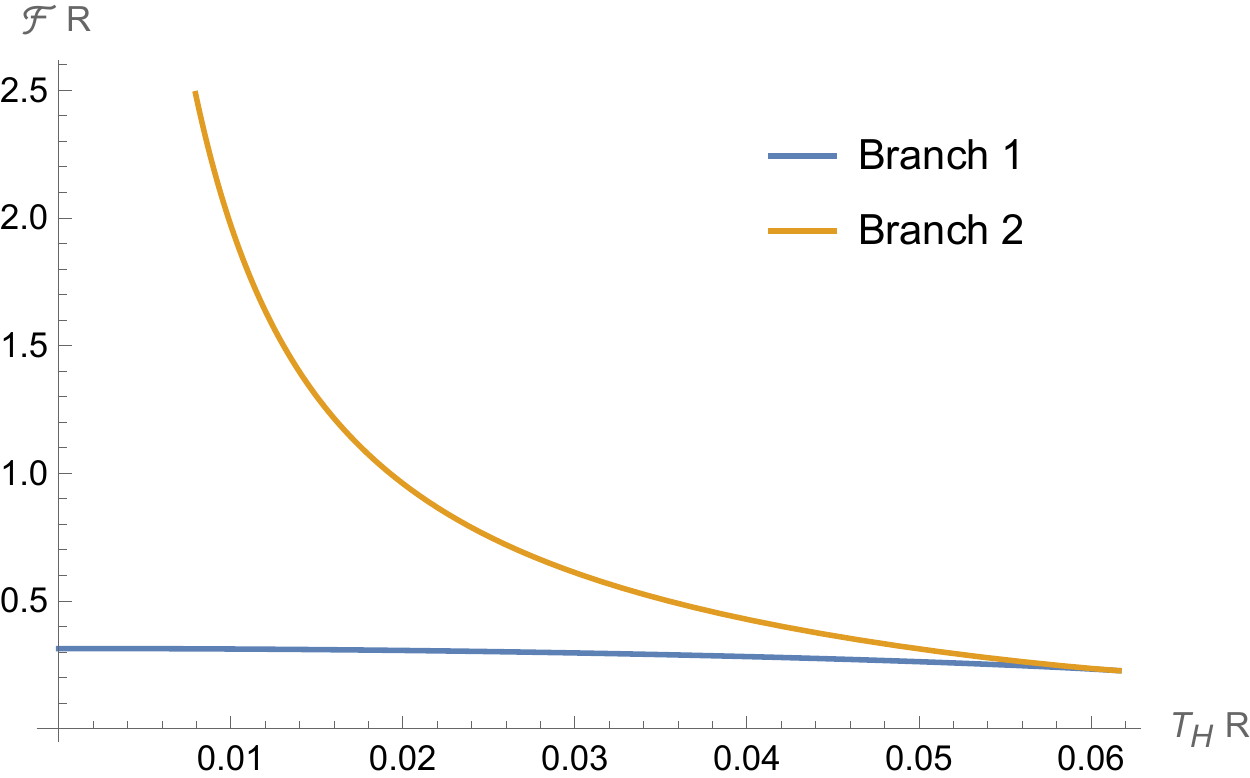}
\caption{Free energy $\mathcal{F}$,  in units of $R^{-1}$, as a function of the temperature, in units of $R^{-1}$, for the two branches of our black-hole model. Branch $I$ and $II$ correspond to solid blue and solid orange curves respectively. We see that ``quantum deformed'' black holes in branch $I$ are always energetically preferred  with respect to  those in branch $II$.}
	\label{PhaseDiagram}
\end{figure}

The existence of this phase transition and  related thermodynamic phase portrait, for the solutions with $0\le R\le R_\text{c}$, can be checked by computing the free energy $\mathcal{F}= M-T_\text{H} S$ as a function of the temperature. The free energy for the two branches $I$ and $II$ has to be calculated numerically by inverting the equation $T_\text{H}=T_\text{H}(\rH)$. We plot  $\mathcal{F}(T_\text{H})$ in \cref{PhaseDiagram}. Branch $I$ corresponds to $\rH$ varying between the minimum value $r=0$ and $r_{\text{H, max}}$. Conversely, branch $II$ corresponds to $\rH$ taking values much larger than $r_{\text{H, max}}$, corresponding to the classical black-hole branch. Notice that the branch $I$ is always energetically preferred with respect to branch $II$.

The presence of a phase transition and a stable branch of small, nonsingular black-hole solution is coherent with the quantum resolution of the classical black-hole singularity. Large  values of $\rH$, i.e., $\rH\gg R$, correspond to the classical, Schwarzschild, thermodynamic  branch, in which black holes are intrinsically unstable. For small values $\rH\sim R$, below the critical temperature $T_\text{H}$, the stable branch of small, quantum, black holes cures the $\rH \to 0$ singular thermodynamic behavior of the Schwarzschild black hole. In the full quantum regime, i.e in the parameter range $R> R_\text{c}$,  black holes do not exist anymore and the gravitational field allows for an effective description in terms of a traversable wormhole.

\subsection{Particle production  and evaporation time}

In this section we will study the Hawking radiation for our regular black hole. We will give a lightning presentation of the derivation, following the original computation in \cite{Hawking:1975vcx}. Since the geometric optics approximation is valid in both cases, the equations of motion will be identical, the only differences being in the metric matching, which will -as we shall show- appear only in the expression of the surface gravity. We start by assuming a Vaidya-like gravitational collapse of a null-shell at the lightcone coordinate $v= v_0$ that leads to the metric \eqref{MetricBrukner}\footnote{A dynamic study of the formation of objects with such metrics from gravitational collapse is an important issue, which is left to future investigations.}. For such a collapse, there is an ``in'' region described by the Minkowski metric
\be
\dd s^2 _{\rm in} = -\dd t^2 +  \dd r^2  +r^2 \dd \Omega^2 ,
\ee 
and an ``out'' region where the metric reads
\be 
\dd s^2 _{\rm out} = -\left[1-\frac{2GM}{r}\Erf \right]\dd t^2 + \frac{\dd r^2}{1-\frac{2GM}{r}\Erf} + \left(r^2 + \frac{3R^2}{4} \right)\dd\Omega^2.
\ee
We consider a massless scalar field $\phi(x)$ obeying the usual Klein-Gordon (KG) equation in the fixed spacetime background given by the previous metric. The field can be expanded in terms of both in and out wave-mode functions 
\be  
\Psi(x) &=  \sum_k a_k u_k(x) + a^\dagger _k u^*_k(x) \\
&= \sum_k b_k v_k(x) + b^\dagger _k v^*_k(x),
\ee
where $a_k$ is the particle annihilation operator in the ``in'' region, $b_k$ in the ``out'' region. $u_k$ and $v_k$ are thus the corresponding ``in'' and ``out'' wave modes.  
Each set of modes is a complete basis and the two sets can be related to each other through the Bogoliubov transformations 
\be \label{eq23}
v_k(x) = \sum_j \alpha_{k j } u_j(x) + \beta_{k j } u_j^* (x) .
\ee
One can then easily check that 
\be \lb{alphabeta}
\alpha_{ k j}  = (v_k , u_j) \quad {\rm and} \quad \beta_{k j } = -(v_k, u_j^*),
\ee
where the canonical inner product $(v_k , u_j)$ is defined by
\be 
\left( v_k(x) ,u_j(x) \right) = \int \dd \Xi \, n^{\mu} \, \left[ v_k(x) \partial_\mu u_j^*(x) -  u_j^*(x) \partial_\mu u_k(x) \right]
\ee
$\Xi$ is a Cauchy hypersurface, and $n^\mu$ its normal vector. This product can be shown to be independent of the choice of the hypersurface (see, e.g., Ref.~\cite{Townsend:1997ku}).

Similarly, also the ``in'' and ``out'' creation and annihilation operators are related through Bogoliobov transformations. We can now compute the expectation value of the number operator of ``out'' modes in the ``in'' vacuum and we find 
\be 
\langle N^{\rm out} \rangle = \sum_j |\beta_{k j}|^2.
\ee
Information about the number of Hawking quanta of each mode $k$ is encoded in the $\beta$ coefficient, which requires evaluating the integral in \eqref{alphabeta}.
To do so, one has to find both the ``in'' and ``out'' wave-modes that are solutions of the KG equation, each in its corresponding spacetime geometry.
Following Hawking's computations \cite{Hawking:1975vcx, Fabbri:2005mw}, one arrives at 
\be\label{ff1} 
\langle N^{\rm out} _\omega \rangle = \sum _{ \omega'} |\beta_{\omega \omega'}|^2 = \frac{1}{e^{- 2 \pi \omega \kappa  } -1 }\, .
\ee
As it is known, the equation for $N^{\rm out}_\omega$ describes a thermal flux of particles with a Planckian spectrum at temperature  $T=\kappa/2\pi$. The only difference with respect to the original Hawking calculation is the explicit expression of the surface gravity $\kappa$ evaluated at the horizon. From the temperature \eqref{Temperaturefinal}, we have
\be
\kappa = \frac{1}{2 \rH} -\frac{\sqrt{\frac{2}{\pi}}\ e^{-2\rH^2/R^2}}{R \ \text{Erf}\left(\frac{\sqrt{2}}{R}\rH \right)}\, .
\ee
\begin{figure}
\centering
\includegraphics[width= 9 cm, height = 9 cm,keepaspectratio]{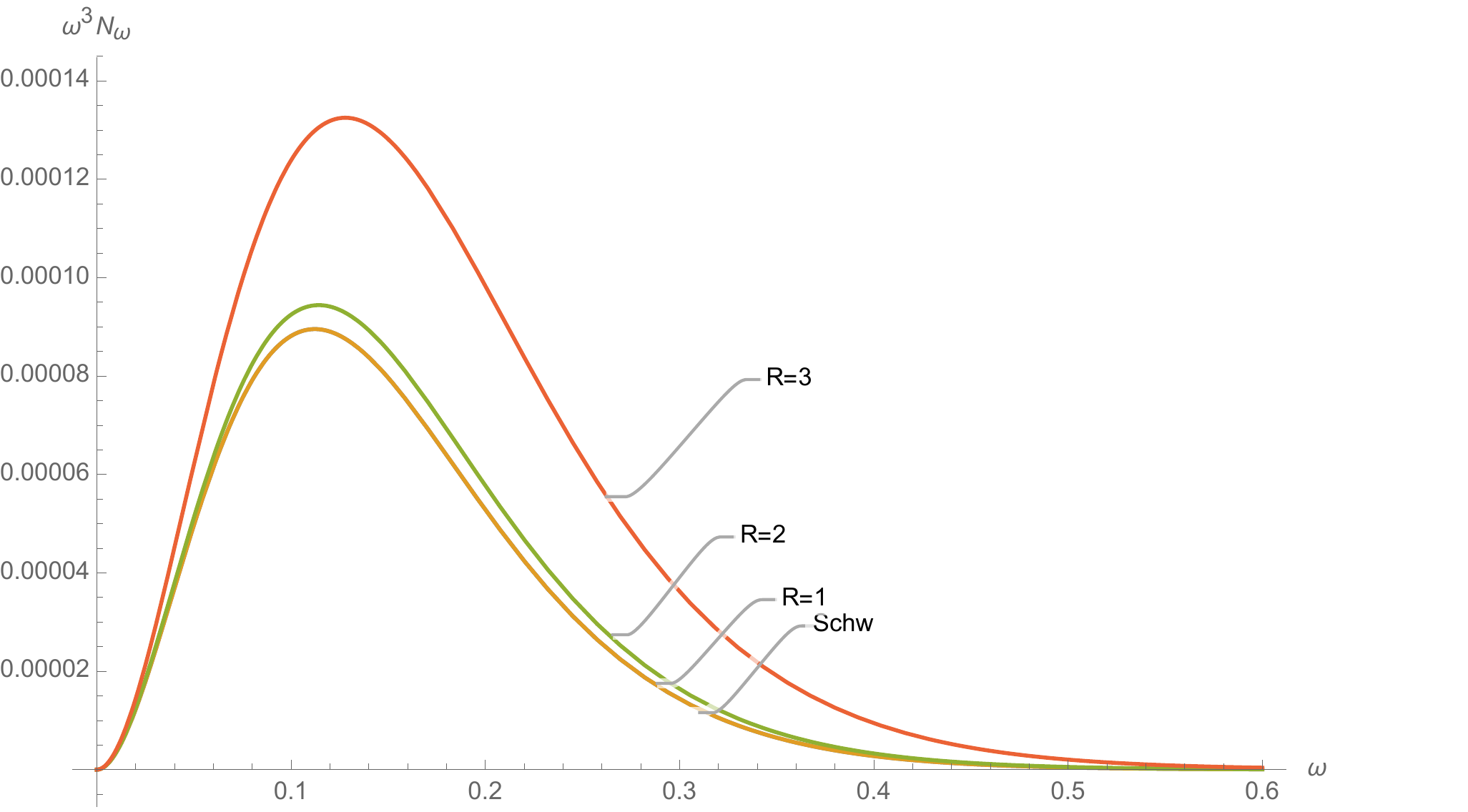}
\caption{Spectral radiance of the black holes for different values of $R$ (in units of $G M$), in comparison with the Schwarzschild case. }
	\label{SpectralRadianceBH}
\end{figure}
Since the spectrum of the quantum corrected black holes is Planckian, we can use the Stefan-Boltzmann law to compute the luminosity,
\be 
L= \sigma_{\rm SB} \mathcal{A}_{\rm H} T_{\rm H}^4,
\ee 
where $\sigma_{\rm SB}$ is the Stefan-Boltzmann constant and $\mathcal{A}_{\rm H} = 4 \pi \left(\rH^2 + 3R^2/4\right)$ is the surface area of the 2-sphere computed at $\rH$. We then use this to compute the mass loss rate which is simply
\be 
\frac{\dd M(t)}{\dd t} = - L = - 4 \pi \, \sigma_{\rm SB} \, \left(\rH^2 + \frac{3}{4}R^2\right) \, T_{\rm H}^4. \lb{Mvar}
\ee
The main problem is the absence of an analytic formula isolating $\rH$ as a function of the black-hole mass $M$. We however get around this difficulty by expressing $M$ in terms of $\rH$ as in \cref{ADMmass}.  
Thus, the variation of $M(t)$ with time is 
\be \label{eq:dMdt}
\frac{\dd M } {\dd t} = \left[ \frac{1}{2 G \text{Erf}\left(\frac{\sqrt{2} \rH(t)}{R}\right)} - \frac{ \sqrt{\frac{2}{\pi }} \rH(t) e^{-\frac{2 \rH(t)^2}{R^2}}} {G \, R \, \text{Erf}\left(\frac{\sqrt{2} \rH(t)}{R}\right)^2} \right] \frac{\dd \rH(t) }{\dd t}\, .
\ee
Plugging \cref{eq:dMdt,Temperaturefinal} into \cref{Mvar} and integrating from an initial radius $r_{\rm H, 0}$ down to $\rH =0$ yields
\be
\Delta t = -\int_{r_{\rm H, 0}}^0 \frac{128 \pi^5 \rH^4 \, R^3\,  e^{6 \rH^2/R^2} \, \text{Erf}\left(\frac{\sqrt{2} \rH}{R}\right)^2 \left[R \, e^{2 \rH^2/R^2} \, \text{Erf}\left(\frac{\sqrt{2} \rH}{R}\right)-2 \sqrt{\frac{2}{\pi }} \, \rH\right]}{G \sigma _{\text{sB}} \left(4 \rH^2+3 R^2\right)  \left(\sqrt{\pi} \, R \, e^{2 \rH^2/R^2} \text{Erf}\left(\frac{\sqrt{2} \rH}{R}\right)-2 \sqrt{2} \, \rH\right)^4} \, \dd \rH.
\ee
Since we are interested in the final part of the evaporation process, i.e., $\rH \sim 0$ (which coincides with the extremal configuration), we look at the expansion of the integrand near this point. We get, after integration, $\Delta t\sim \left[\frac{R^5}{G \, \rH^2} + \mathcal{O}(\rH^{-3})\right]\biggr|_{r_{\text{H},0}}^0$, which diverges in $\rH =0$. This is perfectly consistent with the thermodynamic behavior analyzed in \cref{sec:Thermodynamics}: for values of $\rH$ smaller than that in correspondence with the temperature peak (see \cref{Fig.Temperature}), we have stable remnants, which therefore should not evaporate.

\section{The phenomenology}
\label{sec:Phenomenology}

The aim of this section is to compute phenomenological observables and to compare them to the Schwarzschild case. In fact, despite the simplicity of our derivation, the  presence of the additional parameter $R$ entering the wave function for the source may have observational signatures, which could be tested in the near future by black-hole imaging and GWs observations. In particular, we will analyze the geodesic structure of our spacetime and the QNMs for a scalar perturbations in the eikonal regime.

\subsection{Geodesic structure}
\label{GeodesicsGeneral}
In order to study the goedesics equation, we start by considering the following Lagrangian in the usual $(t,r,\theta,\varphi)$ Schwarzschild coordinates, which can be easily derived from the metric \eqref{MetricBrukner}
\be
    \label{eq:lagrangian}
    \mathscr{L} &= \frac{1}{2}g_{\mu\nu}\xdot^\mu \xdot^\nu = \frac{1}{2}\left[- f(r) \tdot^2 + \frac{\rdot^2}{f(r)}+\left(r^2+\frac{3 R^2}{4}\right)\left(\thetadot^2+\sin ^2\theta\ \phidot^2\right)\right],
\ee
where the dot indicates differentiation with respect to some affine parameter $\lambda$. The equations of motion of a particle in such a spacetime are given by
\begin{equation}
    \label{eq:geodesics_from_lagrangian}
    \left(\frac{\partial}{\partial\lambda}\frac{\partial}{\partial \xdot^\mu}-\frac{\partial}{\partial x^\mu}\right)\mathscr{L}=0,
\end{equation}
and the conjugate momenta are given by
\begin{subequations}\label{eq:conjugate_momenta_geodesics}
\be
    p_t &= \partial_{\tdot}\mathscr{L} = - f(r) \tdot,\quad &p_r &= \partial_{\rdot}\mathscr{L}=\frac{\rdot}{f(r)},\\
    p_\theta &= \partial_{\thetadot} \mathscr{L} = \left(r^2+\frac{3}{4}R^2\right)\thetadot,\quad%
    &p_\varphi &= \partial_{\phidot} \mathscr{L} = \left(r^2+\frac{3}{4}R^2\right)\sin^2\theta\,\phidot.
\ee
\end{subequations}
Notice that the lagrangian is not explicitly dependent on $t$ and $\varphi$. The corresponding quantities $p_t = -E$ and $p_\varphi=L$ are conserved, a clear consequence of the isometries of the metric. Moreover, from the equations of motion, it follows that
\begin{equation}
    \frac{\partial p_\theta}{\partial \lambda}=\frac{\partial}{\partial\lambda}\left(r^2\thetadot\right)=-\frac{\partial\mathscr{L}}{\partial \theta}=\left(r^2+\frac{3}{4}R^2\right)\sin\theta\cos\theta\,\phidot^2,
\end{equation}
so that, if we choose $\theta=\pi/2$ when $\thetadot$ is zero, $\ddot{\theta}$ will be zero as well, and the motion will be constrained on the equatorial plane since $\theta$ will remain constant at the assigned value. In order to find another integral of motion, we can build the hamiltonian corresponding to the lagrangian \eqref{eq:lagrangian} as
\begin{equation}
    \label{eq:hamiltonian}
    \mathcal{H} = p_\mu \xdot^{\mu}-\mathscr{L}. %=\mathscr{L}.
\end{equation}
It is straightforward to see that neither the lagrangian nor the hamiltonian depend on the affine parameter, therefore $\mathcal{H}=\mathscr{L}=\text{const.}=-\epsilon^2/2$, where $\epsilon = 0$ or $\epsilon = \pm 1$ for null and time-like geodesics, respectively. From the constancy of the lagrangian, we can write
\begin{equation}\label{radialequationgeodesic}
    \rdot^{2}+f(r)\left(\epsilon^2+\frac{L^2}{r^2+3R^2/4}\right)=E^2,
\end{equation}
which is the desired equation for geodesics in the spacetime \eqref{MetricBrukner}.

\subsection{Time-like geodesics}

\subsubsection{Proper time of radial infalling time-like particles}

We want to compute the proper-time interval a massive particle in radial free-fall in the metric \eqref{MetricBrukner} takes to reach $r=0$ starting from some finite distance $r=r_0$. We take radial (infalling) time-like geodesics, which therefore satisfy the constraint
\begin{equation}\label{timelikeconstraintradialgeod}
    g_{\mu\nu}u^\mu u^\nu =-1,
\end{equation}
together with $u^\theta = u^\varphi =0$. Using the geodesic integral of motion $\dot t = E/f$,  \cref{timelikeconstraintradialgeod} translates to
\begin{equation}\label{rdotsquare}
    -\frac{E^2}{f(r)}+\frac{\dot{r}^2}{f(r)}=-1 \Rightarrow \dot{r}^2 =E^2 -f(r).
\end{equation}
Since the value of $E$ will not alter the qualitative results of this section, we can choose $E=1$, which means that the particle starts at infinity at rest (marginally bound geodesics). Since we are interested in the behavior near $r=0$, we expand \cref{rdotsquare} around $r=0$
\begin{equation}
    \left(\frac{\dd r}{\dd\tau} \right)^2 \simeq \frac{4GM}{R}\sqrt{\frac{2}{\pi}}-\frac{8GM}{3R^3}\sqrt{\frac{2}{\pi}}r^2 \equiv \frac{R_{\rm c}}{R}-\frac{2R_{\rm c}}{3R^3} r^2
\end{equation}
where we defined $\lambda=\tau$ as the proper time and $R_{\rm c}$ is  the critical value of $R$ at which we have the transition to a horizonless wormhole, i.e., $4\sqrt{2/\pi} GM$. Therefore, the proper time, as measured by a particle moving from $r_0$ to $r$, is given by the integral
\begin{equation}
    \Delta \tau(r) = -\int_{r_0}^r \frac{\dd r'}{\sqrt{\frac{R_{\rm c}}{R}-\frac{2R_{\rm c}}{3R^3}r'^2}}
\end{equation}
where the minus accounts for radial infalling geodesics. This integral is analytical. Evaluating it in the limit $r \to 0$ yields the finite results
\begin{equation}\label{eq:finite_time_for_inf_mass}
    \Delta \tau (r\to 0) = \sqrt{\frac{3}{2}} \frac{R^{3/2}}{\sqrt{R_{\rm c}}}\,  \text{arctg}\left[\frac{\sqrt{2R_{\rm c}}\ r_0}{ \sqrt{R_{\rm c}(-2r_0^2+3R^2)}} \right].
\end{equation}
%
% \begin{figure}
% \centering
% \includegraphics[width= 9 cm, height = 9 cm,keepaspectratio]{Figures/ProperTime.pdf}
% \caption{Result of the numerical integration of the proper time interval \eqref{rdotsquare} required for a proper mass to reach $r=0$, starting from a distance of $r_0= 10 GM$, in the quantum-corrected geometry with $R= 2GM$. We set $G=M=1$. \maurcomment{1) why should this curve be a function of $r$ if it is the time needed to reach $r=0$ and 2) is this really necessary?}}
% 	\label{ProperTime}
% \end{figure}
%
This result is particularly important in the horizonless wormhole case, since it confirms that indeed it is traversable, as massive particles can reach the throat in a finite interval of proper time.

\subsubsection{Time-like geodesic congruence}

We can now study the expansion rate of the metric to check whether the spacetime is geodesically complete or not. To do so, we need to compute the geodesics expansion rate as
\begin{equation}
    \label{eq:expansion_rate}
    \frac{\dd\Theta}{\dd\tau}=\rdot\frac{\dd\Theta}{\dd r}=\rdot\frac{\dd}{\dd r}\left[\frac{1}{\sqrt{-g}}\partial_\mu\left(\sqrt{-g}u^\mu\right)\right],
\end{equation}
where $u^\mu$ is the 4-velocity of a particle orbiting the quantum black hole. If we consider time-like radial geodesics, the components of the vector $u^\mu$ read
\begin{equation}
    u^\mu=\left(\frac{1}{f(r)},\pm\sqrt{1-f(r)},0,0\right),
\end{equation}
where the upper (lower) sign refers to outgoing (ingoing) geodesics. Therefore, \cref{eq:expansion_rate} becomes just
\begin{equation}
    \label{eq:expansion_rate_ur}
    \frac{\dd\Theta}{\dd\tau}=\rdot\, \frac{\dd}{\dd r}\left[\frac{1}{\rho^2}\frac{\dd}{\dd r}\left(\rho^2\,u^r\right)\right],
\end{equation}
where $\rho^2=r^2+3R^2/4$. The evaluation of \cref{eq:expansion_rate_ur} in terms of the metric functions around $r=0$ gives
\begin{equation}
    \frac{\dd\Theta}{\dd\tau}=-4\left(\frac{2}{\pi}\right)^{1/4}\sqrt{\frac{G M}{R^5}}\, \rdot + O(r^2),
\end{equation}
which indicates that the solution is regular near this point and caustics cannot form. 

\subsection{Null geodesics}

\subsubsection{Null-geodesic congruence}
We start from a null vector field $k^\mu =\dd x^\mu/\dd\lambda$ (where $\lambda$ is as usual the affine parameter), satisfying the normalization condition $k_\mu k^\mu =0$, tangent to a bundle of radial in-going null geodesics. The null-geodesic congruence therefore reads
\begin{equation}
    \Theta = \nabla_\mu k^\mu = \frac{1}{\sqrt{-g}}\partial_\mu \left(\sqrt{-g} k^\mu \right).
\end{equation}
We first need to compute the components of the vector field $k^\mu$. To do so, we consider again radial null geodesics, setting $\theta = \text{constant}$ and $\varphi = \text{constant}$.
Also in this case it is useful to introduce the Eddington-Finkelstein coordinates
%From this, if we again use Eddington-Finkelstein coordinates
%
\begin{equation}\label{EFcoordinates}
    u = t-r_\ast, \quad v = t+ r_\ast, \quad r_\ast = \int f^{-1} \dd r.
\end{equation}
We see that the vector field $k_\mu=-\partial_\mu u$ is tangent to the outgoing geodesics, while $k_\mu = -\partial_\mu v$ is tangent to the ingoing ones. We are interested in the latter, whose components are $k_\mu=(-1,-f^{-1},0,0)$. So, we also have $k^\mu = g^{\mu\nu}k_\nu = (f^{-1},-1,0,0)$, and we see that the constraint $k_\mu k^\mu=0$ is satisfied. Therefore, the congruence reads
\begin{equation}
    \Theta = -\frac{1}{r^2+\frac{3R^2}{4}} \partial_r \left(r^2+\frac{3R^2}{4} \right) = -\frac{8r}{4r^2+3R^2}\, .
\end{equation}
The null-geodesic expansion thus reduces to
\begin{equation}
    \frac{\dd\Theta}{\dd\lambda} = \frac{\dd\Theta}{\dd r} k^r = \frac{8 \left(4 r^2-3 R^2\right)}{\left(4 r^2+3 R^2\right)^2},
\end{equation}
which near $r=0$ behaves as
\begin{equation}
    \frac{\dd\Theta}{\dd\lambda}\simeq -\frac{8}{3 R^2}+\frac{32}{3 R^4}r^2 + \mathcal{O}(r^3).
\end{equation}
As in the previous case, no caustics form.
%so once again, no caustics form.

\subsubsection{Photon sphere}
\begin{figure}
    \centering
    \includegraphics[scale=0.4]{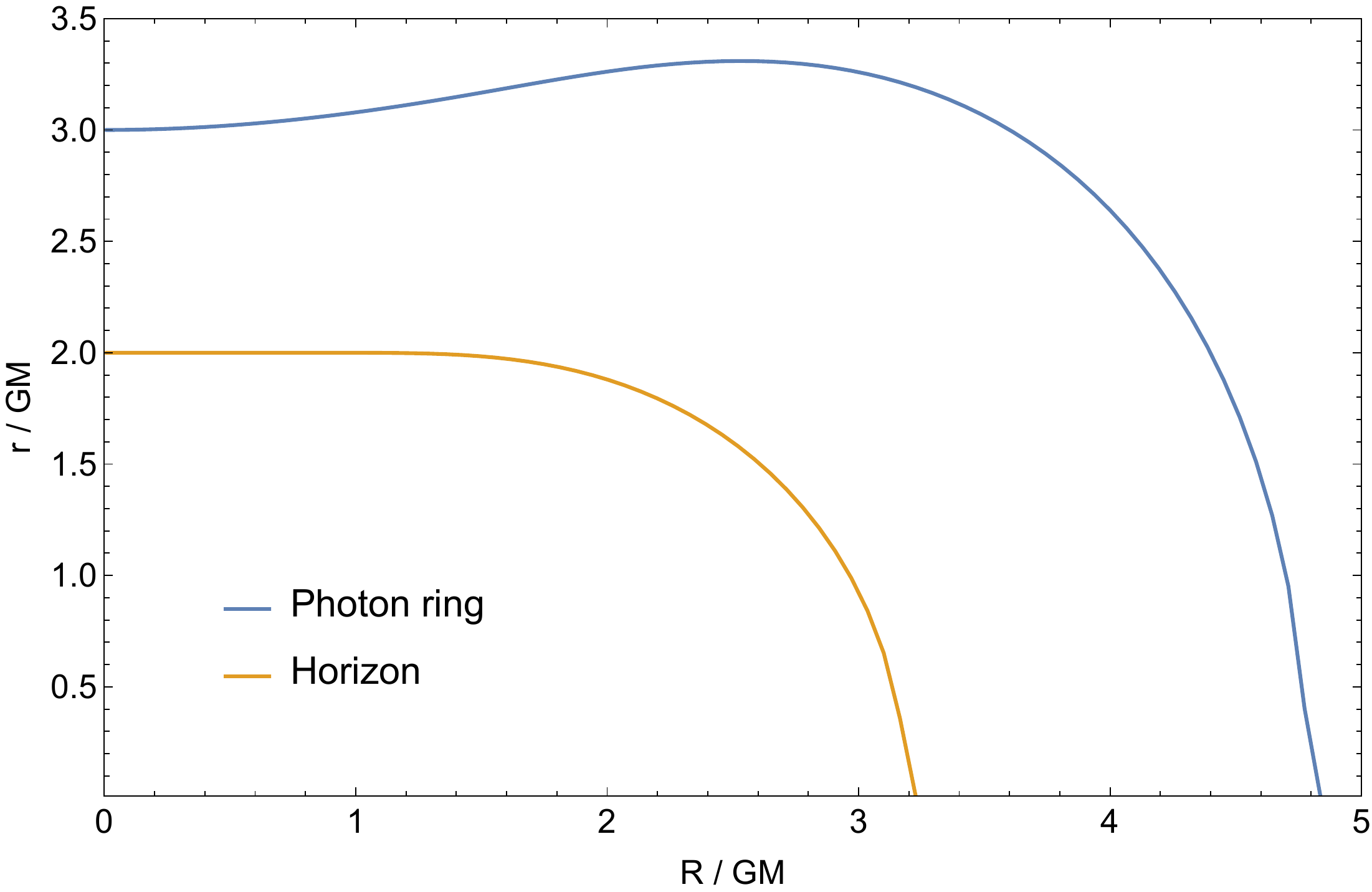}
    \caption{Position of the photon ring and horizon radius as a function of $R$, both in units of $GM$.}
    \label{fig:ph_ring}
\end{figure}

The second term in the left-hand side of \cref{radialequationgeodesic} (for null geodesics), i.e.,
\begin{align}\label{eq:eff_pot_phopton_sphere}
V(r)=f(r)\frac{L^2}{r^2+3R^2/4},
\end{align}
can be thought of as an effective potential felt by massless particles orbiting around either the black hole or the wormhole. Therefore, minima and maxima of this potential correspond to the radii of unstable and stable orbits, respectively. In order to determine  the position of such points, we have to find the zeroes of $\dd V(r)/\dd r$, i.e., the roots of the equation
\begin{align}\label{eq:ph_sphere}
-3 G M R\, e^{\frac{2 r^2}{R^2}} \left(4 r^2+R^2\right) \Erf+2 \sqrt{\frac{2}{\pi }} G M r \left(4 r^2+3 R^2\right)+4 r^3 R \, e^{\frac{2 r^2}{R^2}}=0.
\end{align}
Inspection of \cref{eq:eff_pot_phopton_sphere} shows that the potential has always a maximum for values of $R$ less than a minimum value $R_{\rm min} \simeq 4.8 GM$, while for larger values the maximum shifts to $r=0$. The maximum corresponds to the so-called photon sphere  (or photon ring). The numerical solution of \cref{eq:ph_sphere} is shown in \cref{fig:ph_ring} as a function of $R$,  from which  we see that there could be potentially detectable deviations from the standard Schwarzschild phenomenology. 
%This result shows that we could have potentially detectable deviations from the standard Schwarzschild phenomenology. 

\begin{figure}
\centering
\subfigure[$R< R_{\rm c}$]{\includegraphics[width=6cm]{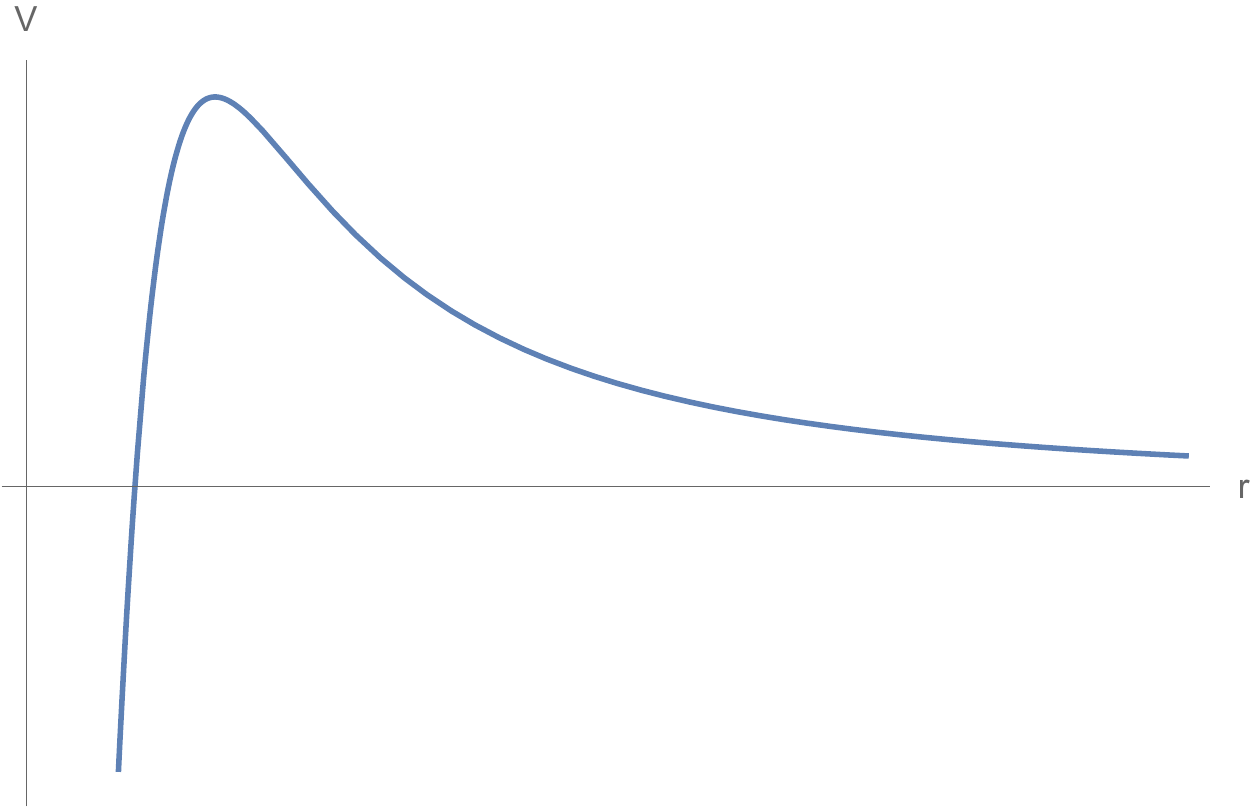}}
\hspace{0.7 cm}
\subfigure[$R = R_{\rm c}$]{\includegraphics[width=6cm]{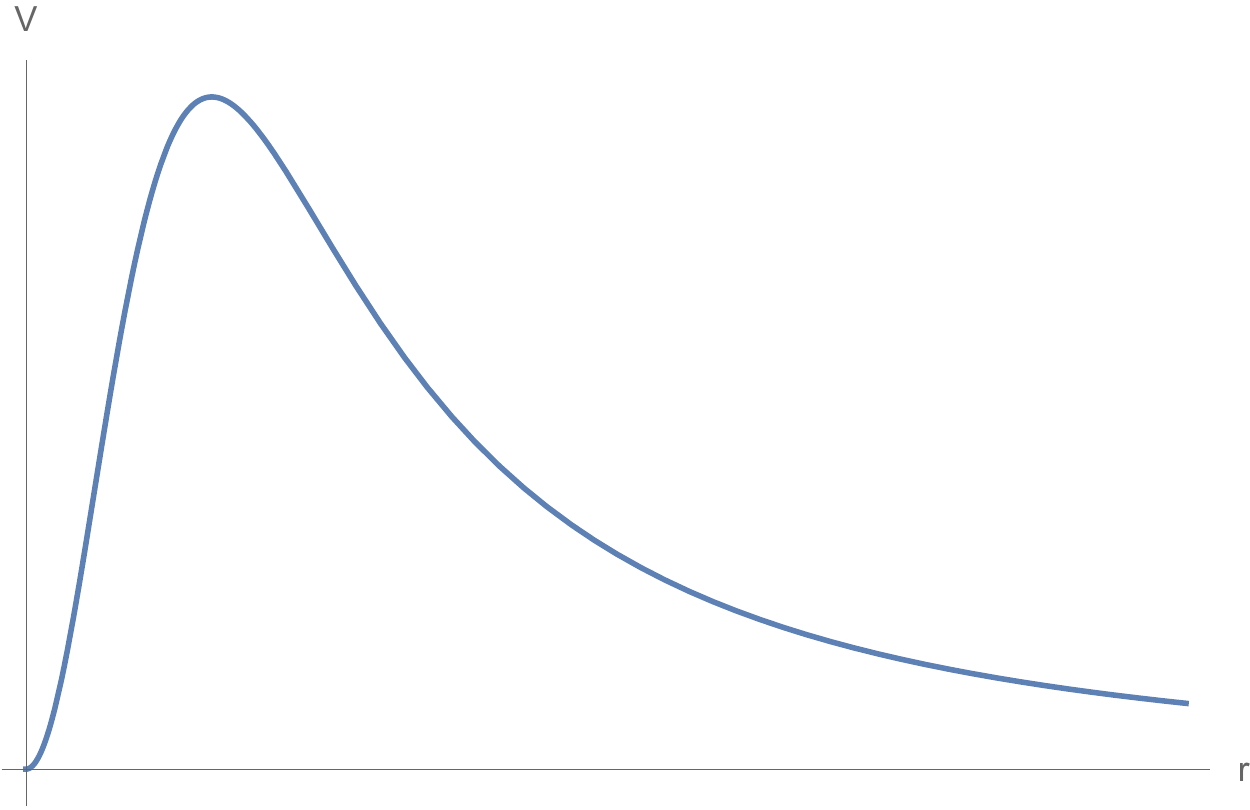}}
\vfill
\centering
\subfigure[$R_{\rm c}<R<R_{\rm min}$]{\includegraphics[width=6cm]{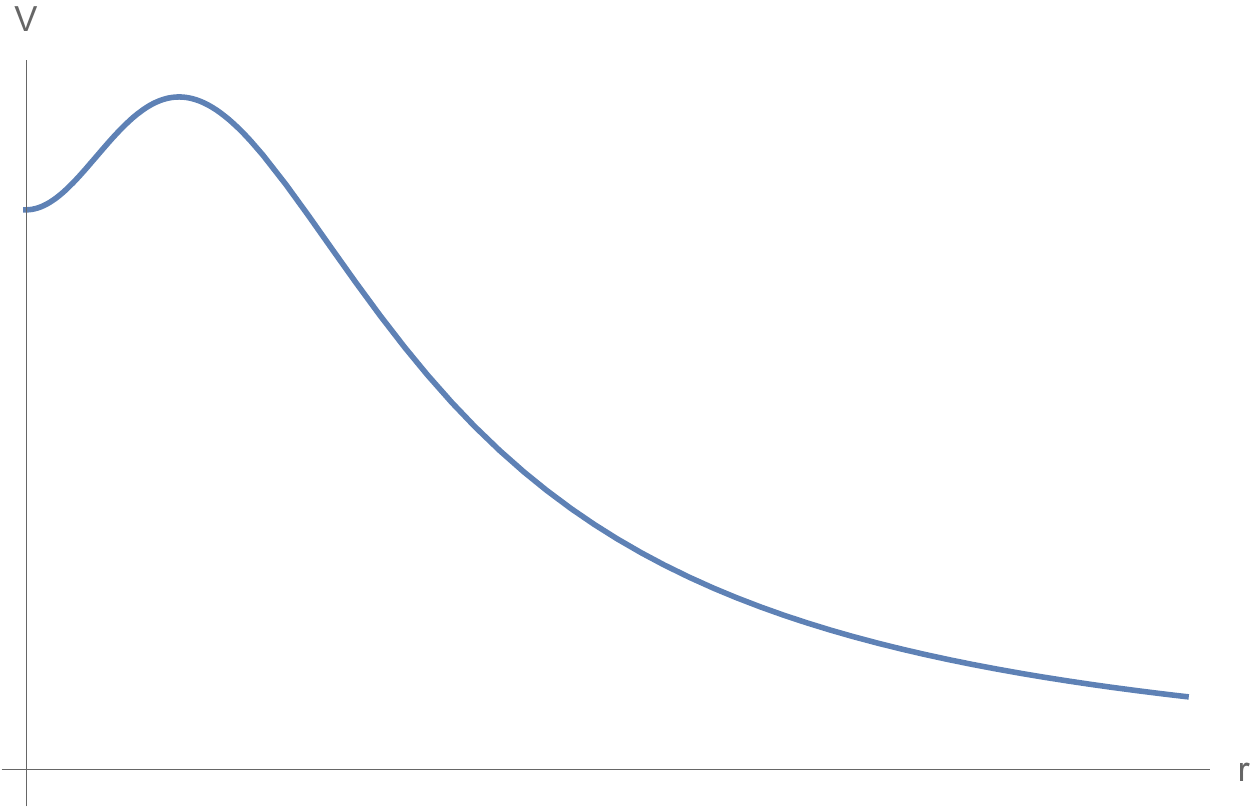}}
\hspace{0.7 cm}
\subfigure[$R>R_{\rm min}$]{\includegraphics[width=6 cm]{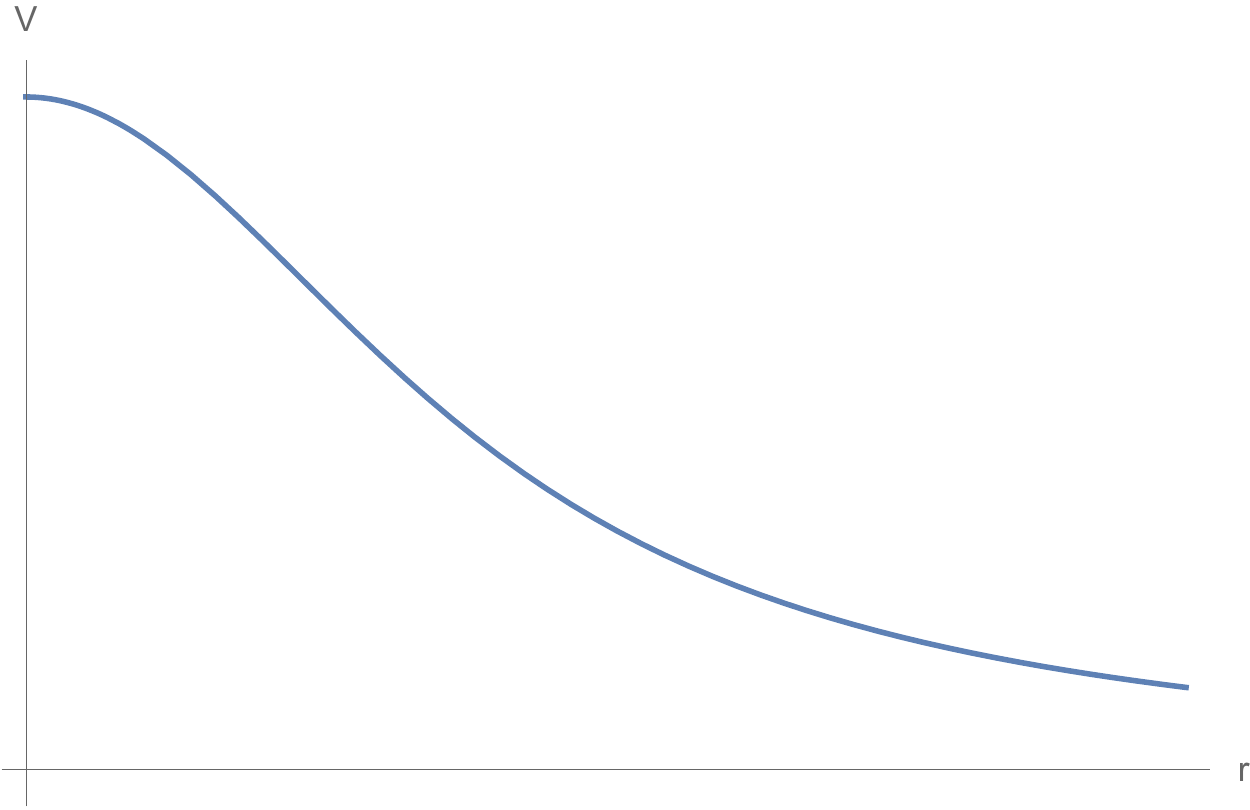}}
\caption{Typical  qualitative  behavior of the effective potential  $V$ for null geodesics as a function of $r$ for $R< R_{\rm c}$,  $R = R_{\rm c}$, $R_{\rm c}<R<R_{\rm min}$ and   
$R>R_{\rm min}$. }
\label{photonOrbgeneral}
\end{figure}

The qualitative behaviors of the effective potential for different values of $R$ is instead depicted in \cref{photonOrbgeneral}. We note that both the ``extremal'' configuration with $R = R_{\rm c}$ and the traversable wormhole with $R_{\rm c}< R < R_{\rm min}$ have also a minimum, corresponding to a  stable photon orbit at $r=0$ (at the throat), which is however excluded. 

\subsection{Scalar perturbations and quasi-normal modes}
In this section we investigate QNMs for scalar perturbations in the fixed background given by our solutions. We will then use the eikonal approximation to give an analytical estimate of the quasi-normal frequencies for the black-hole model.

\begin{figure}
\centering
\subfigure[ $R < R_{\rm c}$]{\includegraphics[width=5cm]{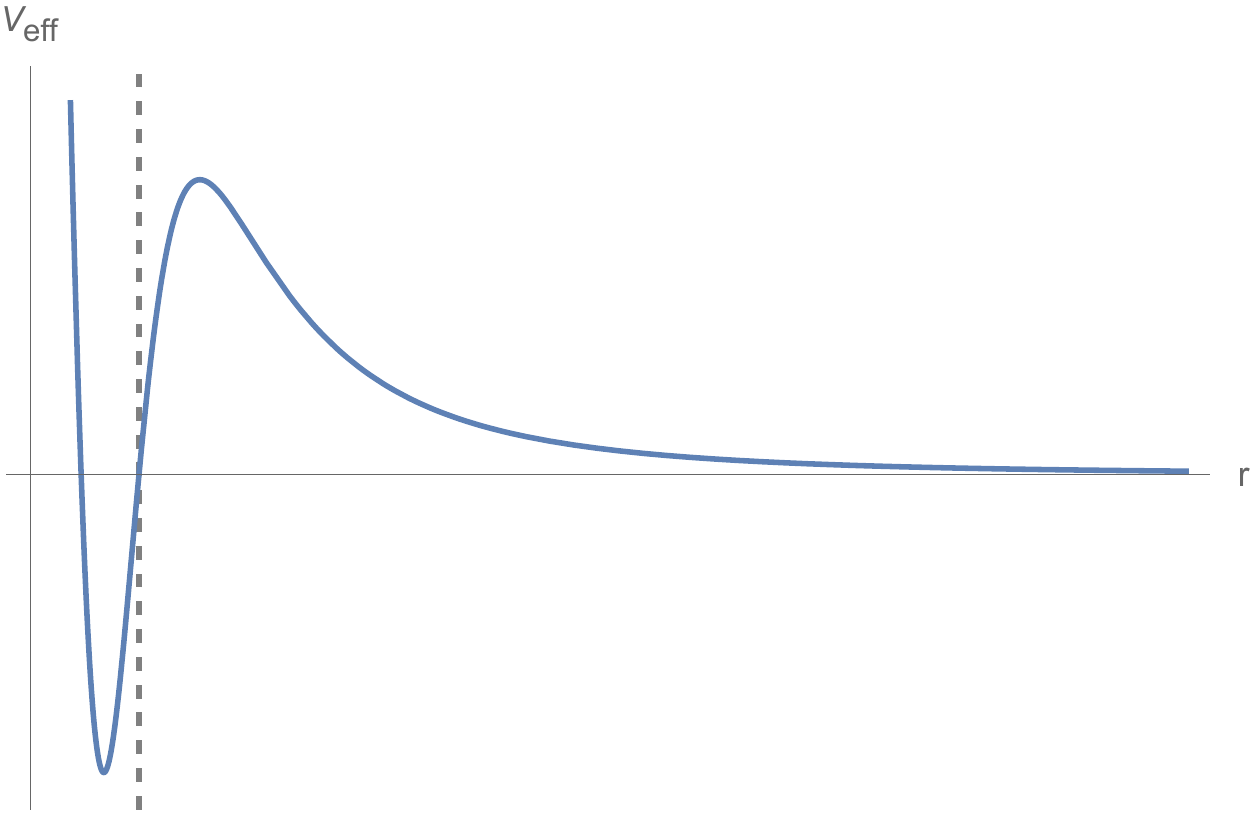}}
\hspace{0.3 cm}
\subfigure[ $R = R_{\rm c}$]{\includegraphics[width=5cm]{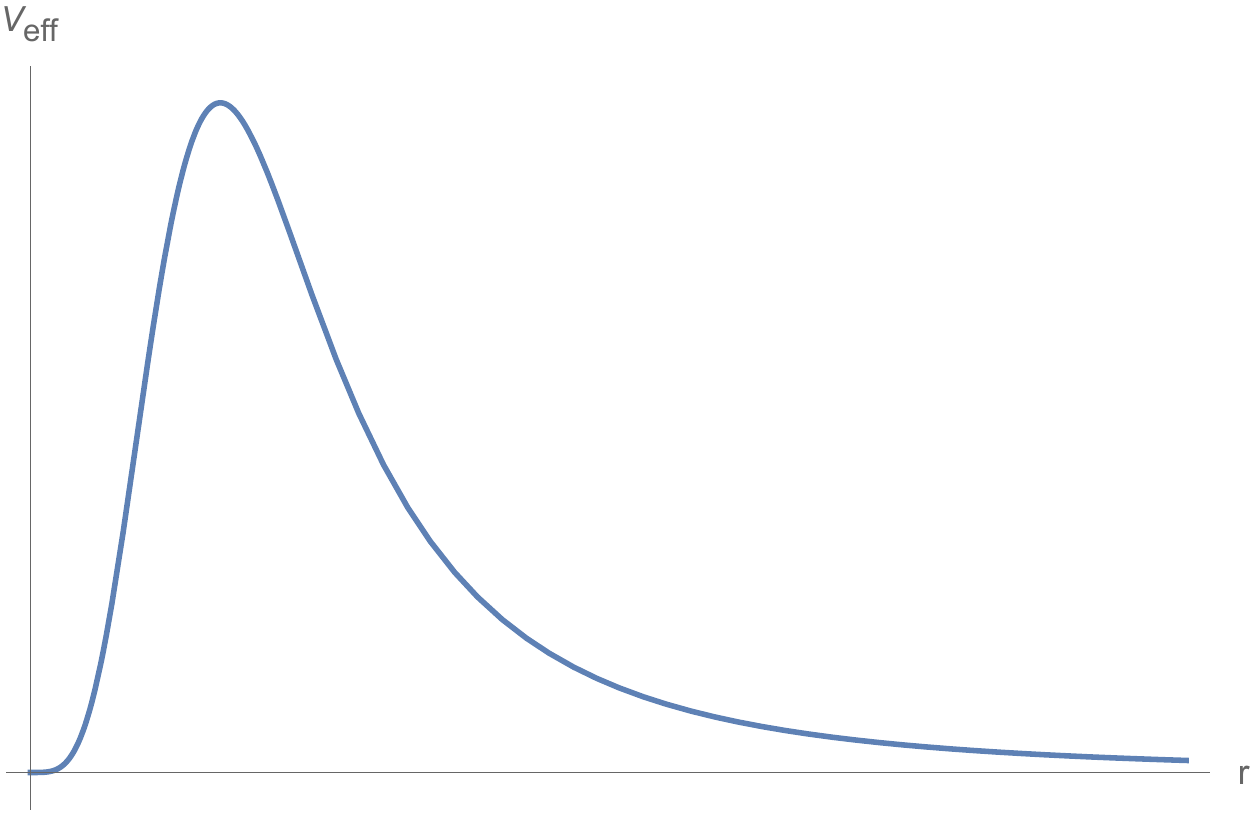}}
\hspace{0.3 cm}
\centering
\subfigure[ $R > R_{\rm c}$]{\includegraphics[width=5cm]{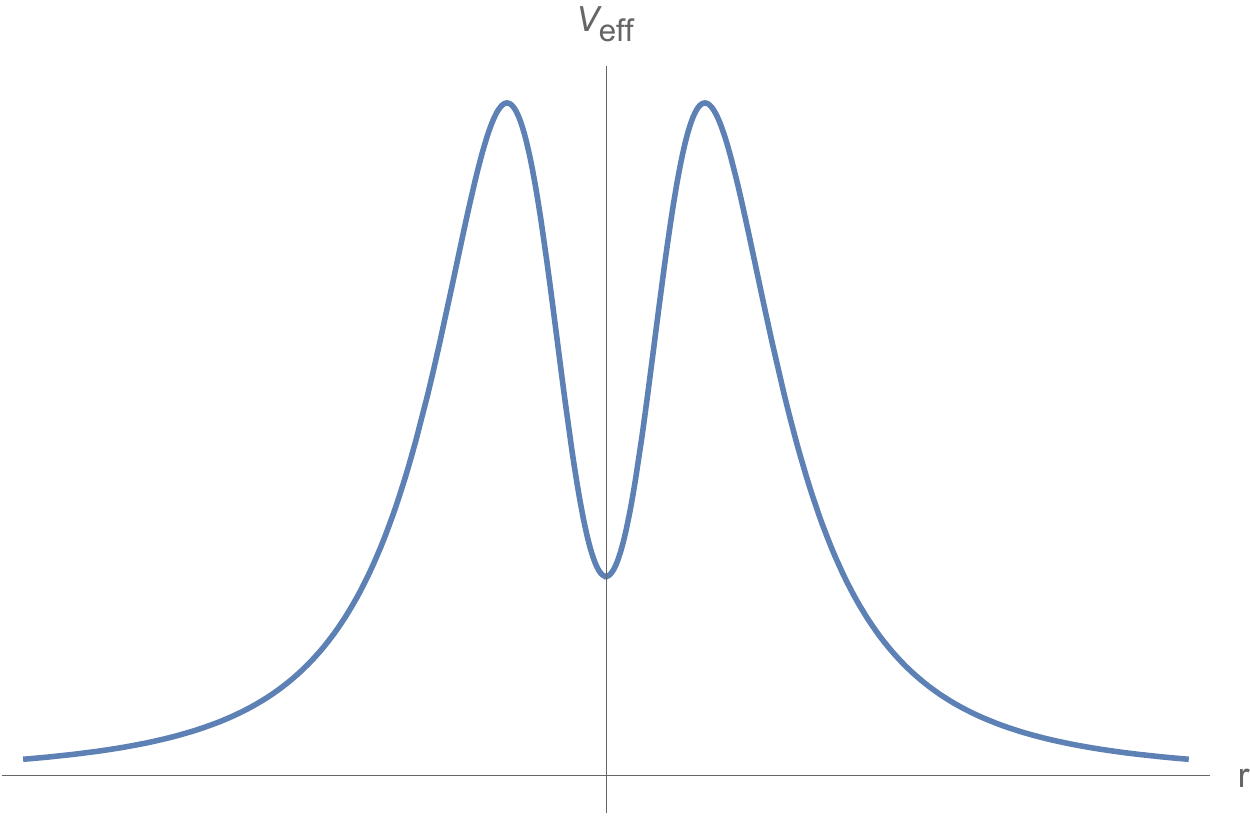}}
\caption{Qualitative behavior of the effective potential \eqref{VeffKG}, with $\ell=0$, for the black-hole model (figure a), the ``extremal model'' (figure b) and the wormhole (figure c). In the first case, the dashed vertical line corresponds to the position of the horizon.}
\label{Veff}
\end{figure}

In order to discuss scalar perturbations and QNMs in our gravitational background, we start from the KG equation for a scalar field in spherical coordinates $\Psi = \Psi(t, r, \theta, \varphi)$ 
\begin{equation}\label{KGequation}
    \Box \Psi = \frac{1}{\sqrt{-g}}\partial_\mu \left(\sqrt{-g} \, g^{\mu\nu} \partial_\nu \right) \Psi= 0
\end{equation}
where $\sqrt{-g}$ is the square root of the determinant of the metric \eqref{MetricBrukner}. Due to the spherical symmetry of the metric, we can separate the angular dependence of $\Psi$ from its radial and temporal dependence, i.e., $\Psi(t,r,\theta,\varphi)\equiv R_{\ell m}(t, r) \mathcal{Y}^{\ell m}(\theta,\varphi)$. The angular part is given in terms of  spherical harmonics, while the radial part satisfies a Schr\"odinger-like equation
\begin{equation}
    \left[\partial_{r_*}^2 - \partial_t^2  -V_\text{eff}(r)\right] \psi =0.
\end{equation}
Here $\psi(t,r)$ is related to $R(t,r)$ by 
\begin{equation}
    R_{\ell m}(t,r) \equiv \frac{\psi_{\ell m}(t,r)}{\sqrt{r^2+\frac{3R^2}{4}}},
\end{equation}
while $V_\text{eff}(r)$ is the effective potential, namely
\begin{equation}\label{VeffKG}
    V_\text{eff}(r) = \frac{12R^2 f^2}{\left(4r^2 + 3R^2 \right)^2} + \frac{4f}{4r^2 + 3R^2}\left[\ell(\ell+1) + rf' \right],
\end{equation}
where primes indicate derivation with respect to $r$.
We see that the presence of both $R$ and a non-trivial angular metric function introduce an additional term in the effective potential with respect to the Schwarzschild case, which goes to zero as $R^2$ in the limit $R\to0$. In \cref{Veff}, we plot some examples for different values of $R$.
The traversable-wormhole case is particularly interesting, as the double peak in the potential always signals the possibility of having echoes in the QNMs spectrum \cite{Cardoso:2016rao,Cardoso:2016oxy,Abedi:2016hgu,Maggio:2020jml,Maggio:2021ans, Chakraborty:2022zlq}.

\subsection{Analytic expression of QNMs in the eikonal limit}
\label{QNMsfrequency}

We can exploit the construction of Ref.~\cite{Cardoso:2008bp} to find an analytic expression of the quasi-normal frequencies in the eikonal regime, i.e., in the $\ell \gg 1$ limit. This construction only works in the case in which the effective potential in the KG equation has a \textit{single} peak (the presence of the double peak in the horizonless-wormhole case invalidates a direct application of this algorithm \cite{Churilova:2019cyt}). The basics of the construction of Ref.~\cite{Cardoso:2008bp} is to exploit a relation between the ringing modes of black holes and photons on the unstable light ring. Specifically, the black-hole vibration modes, whose energy is gradually being radiated away, are interpreted as photons moving along an unstable null-geodesics and slowly leaking out. The real part of the quasi-normal spectrum (corresponding to the periodic oscillations of the modes) is given by the angular velocity of photons on the light ring, whose position is at $\rLR$, namely $\Omega = \phidot/\dot t|_{r=\rLR}$. The imaginary part of the quasi-normal frequencies, instead, responsible for the damping of the modes, is associated to the time-scale of the instability of the circular null geodesics, given by the Lyapunov exponent, whose form reads
\begin{equation}\label{Lyapunovexponent}
    \lambda = \sqrt{-\frac{V''(r)}{2\dot t^2}}\biggr|_{r=\rLR},
\end{equation}
where $V(r)$ is the potential for null geodesics \eqref{eq:eff_pot_phopton_sphere}, while the minus sign is required since the light ring corresponds to an unstable orbit.
Therefore, the spectrum of QNMs reads, in the eikonal regime
\begin{equation}\label{omegaQNMs}
    \omega_{\rm QNMs} = \Omega \ \ell - \ii \left(n+\frac{1}{2} \right)\lambda
\end{equation}
with $n$ an integer (the overtone number). $\Omega$ can be easily computed exploiting \cref{eq:conjugate_momenta_geodesics} and the fact that $p_t=E$ and $p_\phi=L$.
Moreover, using \cref{radialequationgeodesic}, setting $\epsilon=0$ to consider null geodesics, and making use of the fact that $\rdot=0$ at the light ring, yield a relation between $E$ and $L$
\begin{equation}\label{ELlightring}
    \frac{E}{L}= \pm \sqrt{\frac{f(\rLR)}{\rLR^2 + \frac{3R^2}{4}}}\, .
\end{equation}
Therefore, $\Omega$ reads
\begin{equation}\label{Omegageneral}
    \Omega = \frac{\phidot}{\dot t}\biggl|_{r=\rLR} = \sqrt{\frac{f(\rLR)}{\rLR^2+\frac{3R^2}{4}}} = 2 \, \sqrt{\frac{f(\rLR)}{4 \rLR^2 + 3R^2}}\, ,
\end{equation}
%
% From \cref{Lyapunovexponent} we have instead
% %
% \begin{equation}
%     \lambda^2 = -\frac{V''(\rLR)}{2 E^2}f^2(\rLR) = -\frac{1}{2}\frac{V''(\rLR) f(\rLR)}{L^2}\left(\rLR^2 + \frac{3R^2}{4} \right),
% \end{equation}
% %
% where in the second equality we have exploited \cref{ELlightring}. 

% From \cref{eq:eff_pot_phopton_sphere}, we also have that the position of the light ring is given by
% %
% \begin{equation}
%     V'(\rLR)=0 \Rightarrow f'(\rLR) = \frac{8 \ \rLR \ f(\rLR)}{4\rLR^2 + 3R^2},
% \end{equation}
% %
% which greatly simplifies $V''(\rLR)$ to 
% %
% \begin{equation}
%     V''(\rLR) = \frac{4 L^2 \left[-8\ f(\rLR) + f''(\rLR) \left(4\rLR^2 + 3R^2 \right) \right]}{\left(4\rLR^2 + 3R^2 \right)^2}\, .
% \end{equation}
To compute $\lambda$, we start from \cref{Lyapunovexponent}. Using the fact that $V'(\rLR)=0$, we simplify the expression for $V''(\rLR)$ and we get
%
% Therefore, this yields
%
\begin{equation}
    \lambda = \sqrt{-\frac{f(r) \left[-8f(r) + f''(r) \left(4r^2 + 3R^2 \right) \right]}{8r^2+6R^2}}\biggl|_{r=\rLR}.
\end{equation}

\begin{table}[]
    \centering
    \begin{tabular}{ccc}
        \toprule[1pt]
        \addlinespace[1pt]
         $R/GM$     &  $GM\Omega$  & $GM\lambda$ \\\addlinespace[1pt]
        \midrule[0.5pt]
        \addlinespace[1.5pt]
         $10^{-4}$  &  0.19245  &  0.19245 
        \\
         $0.5$  & 0.19048   &  0.19179 
        \\
         $1$  &  0.18508  & 0.18978  
        \\
         $1.2$  & 0.18219   &  0.18861 
        \\
         $1.4$   & 0.17901    & 0.18716   
         \\
          $1.6$   & 0.17562    & 0.18511   
        \\
          $1.8$   &  0.17211   & 0.18201   
        \\
          $2$   & 0.16856   &  0.17754  
        \\
          $2.2$   & 0.16506   & 0.17171   
        \\
          $2.4$   & 0.16167   & 0.16470   
        \\
          $2.6$   & 0.15845   & 0.1567   
        \\
          $2.8$   & 0.15542   & 0.14803   
        \\
          $3$   & 0.15262   & 0.13877   
        \\
          $R_{\rm c}$   & 0.15015   & 0.12945   
        \\\addlinespace[1.5pt]
        \bottomrule[1pt]
    \end{tabular}
    \caption{\label{table1}
    Values of $\Omega$ and $\lambda$, which determine the QNMs frequencies in the eikonal limit through \cref{omegaQNMs}, for different values of the quantum-deformation parameter $R$ (in units of $GM$).
    }
\end{table}

By numerically solving \cref{eq:ph_sphere} to find the position of the light ring for different values of $R$ (limited to the black-hole and the ``extremal'' model cases), one can find the explicit values of the quasi-normal frequencies \eqref{omegaQNMs}, given the values of $\Omega$ and $\lambda$ reported in \cref{table1}. We also checked that, in the small $R$ limit (the first value in \cref{table1}), the quasi-normal frequencies are consistent with the Schwarzschild ones in the eikonal regime, for which $\Omega = |\lambda| = \frac{1}{3\sqrt{3}\, GM}$ \cite{blome1984quasi}. 

Finally, we can study how QNMs behave near the extremal configuration. We can expand both $\Omega$ and $\lambda$ around $R = R_{\rm c}$ (before computing them at $\rLR$). We get
\begin{subequations}
\be
    &\Omega \simeq a + b \ \left(R-R_{\rm c} \right)\sim a + b' \ (M-M_{\rm c});\\
    &\lambda \simeq d + e \ (R-R_{\rm c}) \sim d + e' \ (M-M_{\rm c}),
\ee
\end{subequations}
where we have defined $a \equiv \Omega\left(R_{\rm c} \right)$, $b \equiv \frac{\dd \Omega}{\dd R}\biggr|_{R = R_{\rm c}}$, $d \equiv \lambda(R_{\rm c})$, $e \equiv \frac{\dd \lambda}{\dd R}\biggr|_{R = R_{\rm c}}$\. If we take the near-horizon \footnote{A problem of the near-horizon limit is that the minimum of the null-geodesic effective potential gets shifted inside the event horizon at soon as we move away from ``extremality''.} limit together with the near-extremal limit, it is easy to see that both the constant and the linear term $R-R_{\rm c}$ of $\lambda^2$ go to zero, and therefore we are left with
\begin{equation}
    \omega_{\rm I} \propto \lambda \propto M-M_{\rm c} \propto T_{\rm H},
\end{equation}
where we took cognisance of \cref{Massgapaboveextremality}. This scaling of the imaginary part of the quasi-normal frequencies with the temperature is consistent with some conjectures \cite{Hod:2008se,Hod:2011zzd,Hod:2012zzb,Zimmerman:2015trm,Joykutty:2021fgj}.%
%\footnote{A problem of the near-horizon limit is that the minimum of the null-geodesic effective potential gets shifted inside the event horizon at soon as we move away from ``extremality''.}
These zero-damped (or nearly zero-damped) modes \cite{Joykutty:2021fgj} would therefore represent a clear
phenomenological signature of the extremal configuration.

\section{Conclusions}
\label{Conclusions}

In this paper we have derived the effective quantum spacetime metric generated by  gravitational sources in quantum superposition of different locations. We have considered a simple case of a Gaussian wave-packet in which the width $R$ represents the uncertainty in the position of the source. The resulting spacetime solution has several distinguishing features, inherited from the ``quantumness'' of the source. Firstly, the uncertainty $\Delta r\sim \lambda_\text{DB}\sim R$ in the position of the source prevents the radius of the transverse two-sphere from shrinking to zero, avoiding the presence of the classical $r=0$ singularity of the Schwarzschild solution. Secondly, $L^2$-integrability implies that all spacetime metrics arising from this superposition are asymptotically flat and indistinguishable from the Schwarzschild solution in the asymptotic $r \to \infty$ region.

On the other hand, the possible models derived from our approach have a different inner-core behavior depending on the strength of the quantum superposition effects, i.e on the comparison between values of $\lambda_\text{DB}\sim R$ and the classical gravitational radius $R_\text{S}=2GM$. In the classical limit $R \ll R_\text{S}$, when the source is classically localized at $r=0$, an event horizon appears to shield the classical singularity at $r=0$. For $R \sim R_\text{S}$, quantum effects begin to become relevant, the horizon is still present, but the singularity at $r=0$ is removed by the quantum uncertainty in the position of the source and a ``quantum hair'' $R$ appears. Finally, when the quantum effects become fully dominant, for $R>R_\text{c} = 4 \sqrt{2/\pi}\, GM$, the horizon disappears and the effective spacetime solution becomes a traversable wormhole.

The above description is fully  consistent  with recent ideas, such as the $\text{ER} = \text{EPR}$ conjecture, which consider the spacetime structure as emerging from quantum entanglement. Moreover, it also explains the cosmic censorship conjecture and the no-hair theorem as emerging features of the classical limit of our models. 

The previous description has been confirmed by our investigation of the thermodynamic properties of the black-hole solutions we have found. Our  quantum characterization of the gravitational  source cures the singular thermodynamic  behavior of the Schwarzschild black hole, i.e., negative specific heat and its divergence at $\rH \to 0$. These problems are solved by the phase transition we have found and by the presence of a stable branch of small nonsingular black-hole solutions with positive specific heat.

We have also shown that the presence of the quantum hair $R$ induces modifications, with respect to the Schwarzschild black hole, in the photon orbits, in the spectrum of QNMs for scalar perturbations and in GW signal (presence of echoes, not explicitly computed here). These deviations are potentially detectable in the near future by third-generation gravitational-wave detectors and by black-hole imaging techniques.

On the other hand, the strength of our approach---the independence of our prediction from the details of the underlying microscopic quantum gravity theory---represents also its main limitation. The lack of knowledge about  the microscopic QG theory is reflected in the intrinsic impossibility, within our model, to determine the probability distribution function $\phi(r)$ from some dynamical equation. We have circumvented this problem by working in the framework of an effective theory, generically described by GR sourced by an anisotropic fluid, whose energy density and equation of state are implicitly determined by $\phi(r)$. However, progress in this direction hinges crucially on our ability to incorporate some details of the microscopic QG theory in our effective description.

\section*{Acknowledgements}
We thank Remo Garattini for useful discussions. We are also indebted to Yi Wang, Haitham Zaraket, and Xi Tong for illuminating discussions at the early stage of the project. Ali Akil wishes to thank the Department of Physics at the University of Cagliari for the hospitality during the period of time when part of this work was done.

\begin{appendix}

\section{Curvature invariants}
\label{sec:curvatureinv}

Computing the Ricci tensor for the metric \cref{MetricBrukner}, one easily finds the Ricci scalar
\begin{equation}
    \mathcal{R} = -\frac{4 e^{-\frac{2 r^2}{R^2}} \left[-3 G M R^5 e^{\frac{2 r^2}{R^2}} \left(3 R^2+8 r^2\right) \Erf +2 \sqrt{\frac{2}{\pi }} G M r \left(9 R^6+18 R^4 r^2+32 R^2 r^4+32 r^6\right)+6 R^5 r^3 e^{\frac{2 r^2}{R^2}}\right]}{R^3 r^3 \left(3 R^2+4 r^2\right)^2}.
\end{equation}
At $r=0$, we have
\begin{equation}
    \mathcal{R}(r=0)= \frac{8 \left(6 \sqrt{\frac{2}{\pi }} G M-R\right)}{3 R^3}\, ,
\end{equation}
which shows no divergences.

We have also computed the other curvature invariants, $\mathcal R_{\mu\nu}\mathcal R^{\mu\nu}$, $\mathcal R_{\mu\nu\rho\sigma}\mathcal R^{\mu\nu\rho\sigma}$ (the Kretschmann scalar) and the Weyl contraction $C_{\mu\nu\rho\sigma}C^{\mu\nu\rho\sigma}= \mathcal R_{\mu\nu\rho\sigma}\mathcal R^{\mu\nu\rho\sigma}-2\mathcal R_{\mu\nu}\mathcal R^{\mu\nu}+\mathcal{R}^2/3$.  \\

The first one reads
\begin{equation}
\begin{split}
    \mathcal R_{\mu\nu}\mathcal R^{\mu\nu} =& \frac{8 e^{-\frac{4 r^2}{R^2}}}{\pi  r^6 R^6 \left(4 r^2+3 R^2\right)^4} \biggl\{8 r^2  \biggl[G^2 M^2 \left(4 r^2+3 R^2\right)^2 \left(64 r^8+96 r^6 R^2+100 r^4 R^4+36 r^2 R^6+9 R^8\right)+\\
    &+3 \sqrt{2 \pi } G M r^2 R^5 e^{\frac{2 r^2}{R^2}} \left(32 r^6+48 r^4 R^2+30 r^2 R^4+9 R^6\right)+9 \pi  r^4 R^{10} e^{\frac{4 r^2}{R^2}} \biggr]-12 G M r R^5 e^{\frac{2 r^2}{R^2}}\Erf \times \\
    & \times \biggl[\sqrt{2 \pi } G M \left(256 r^8+480 r^6 R^2+384 r^4 R^4+162 r^2 R^6+27 R^8\right)+9 \pi  r^2 R^5 e^{\frac{2 r^2}{R^2}} \left(4 r^2+R^2\right) \biggr]+\\
    &+9 \pi  G^2 M^2 R^{10} e^{\frac{4 r^2}{R^2}} \left(80 r^4+48 r^2 R^2+9 R^4\right) \Erf^2
    \biggr\}\, ,
\end{split}
\end{equation}
which, evaluated at $r=0$ gives
\begin{equation}
     \left.\mathcal R_{\mu\nu}\mathcal R^{\mu\nu}\right|_{r=0}=\frac{64 \left(36 G^2 M^2-6 \sqrt{2 \pi } G M R+\pi  R^2\right)}{9 \pi  R^6}\, .
\end{equation}
The Kretschmann scalar instead reads
\begin{equation}
\begin{split}
    \mathcal R_{\mu\nu\rho\sigma}\mathcal R^{\mu\nu\rho\sigma}=&\frac{16 e^{-\frac{4 r^2}{R^2}}}{\pi  r^6 R^6 \left(4 r^2+3 R^2\right)^4} \biggl\{8 G^2 M^2 r^2 \left(4 r^2+3 R^2\right)^2 \left(64 r^8+160 r^6 R^2+164 r^4 R^4+60 r^2 R^6+9 R^8\right)+\\
    &-96 \sqrt{2 \pi } G M r^6 R^7 e^{\frac{2 r^2}{R^2}} \left(4 r^2+3 R^2\right)+108 \pi  r^6 R^{10} e^{\frac{4 r^2}{R^2}}-4 G M r R^3 e^{\frac{2 r^2}{R^2}} \Erf \times \\
    &\times \biggl[\sqrt{2 \pi } G M \left(512 r^{10}+2048 r^8 R^2+2688 r^6 R^4+1728 r^4 R^6+594 r^2 R^8+81 R^{10}\right)+\\
    &+12 \pi  r^4 R^5 e^{\frac{2 r^2}{R^2}} \left(3 R^2-8 r^2\right) \biggr]+3 \pi  G^2 M^2 R^6 e^{\frac{4 r^2}{R^2}} \Erf^2\times \\
    &\times \left(256 r^8+256 r^6 R^2+336 r^4 R^4+144 r^2 R^6+27 R^8\right)\biggr\},
\end{split}
\end{equation}
which at $r=0$ reduces to
\begin{equation}
     \left.\mathcal R_{\mu\nu\rho\sigma}\mathcal R^{\mu\nu\rho\sigma}\right|_{r=0}=\frac{64 \left(72 G^2 M^2-16 \sqrt{2 \pi } G M R+3 \pi  R^2\right)}{9 \pi  R^6}. 
\end{equation}
We see that $\mathcal{R}$, $ \mathcal R_{\mu\nu}\mathcal R^{\mu\nu}$ and $\mathcal R_{\mu\nu\rho\sigma}\mathcal R^{\mu\nu\rho\sigma}$ are all regular at $r=0$. This  is a sufficient condition to have also a regular Weyl contraction in this point.

\end{appendix}

\bibliography{refs}
\end{document}